# Eclipse-minimising sun-synchronous orbits in a mean-Sun rotating frame

How a satellite can stay out of Earth's shadow all year — why the usual dawn–dusk orbit falls just short, at what altitude it stops falling short, and how much margin you have

Kyle Cooper · Yanni Nikas

Interactive model: colure.space/kelvinzero ↗

ABSTRACT

A satellite in low Earth orbit normally passes through Earth's shadow on every revolution. A dawn–dusk sun-synchronous orbit largely avoids it: J2 precession keeps the orbit plane facing the Sun all year — but it follows only the Sun's average position, the mean Sun. In reality the Sun wanders up to 23° around this average, enough to push the plane into shadow for part of the year. We introduce MSSI, a coordinate frame that rotates with the mean Sun. In it, a sun-synchronous orbit is a fixed ring and the real Sun traces the analemma, so the yearly eclipse question becomes static geometry. From that picture we derive how much of each orbit is in shadow, the yearly eclipse for every plane orientation, maps of which directions are never, sometimes or always eclipsed, and the margin a mission has. Fixed planes become eclipse-free above about 1400 km, within a window of node offsets that widens with altitude, and the 06:00 and 18:00 lines turn out to be equivalent. The closed form is checked against a full-force-model numerical propagation and against the published eclipse seasons of three flying missions; the model, its code and an interactive 3-D visualizer are published as open source.

**Keywords** sun-synchronous orbit · eclipse · dawn–dusk orbit · J2 nodal precession · analemma · equation of time · mean-Sun frame

## 1 Why an orbit that never sees shadow

A satellite in low Earth orbit circles the planet every 90 to 110 minutes and, on most orbits, passes through Earth's shadow on every lap, for up to about 35 minutes. Each pass is a small crisis the spacecraft has been designed around: the solar arrays stop producing power, so batteries must carry every load; the temperature of exposed hardware drops by tens of degrees within minutes, so heaters switch on; and when the Sun returns everything warms again. Over a five-year mission that is roughly 25,000 hot–cold cycles. Batteries, solder joints, optics and mechanisms all age with cycling, and a large share of a small satellite's mass is there only to survive the dark half of every orbit.

An orbit that never entered the shadow would remove much of that. Power would be continuous, so the battery could be sized as a buffer rather than for every orbit's night; temperatures would settle toward a steady state, reducing heater power and simplifying the thermal design; and payloads that depend on thermal stability — telescopes, lasers, precision clocks — would get a steadier environment. **Project**

**Kelvin Zero** is an effort to find out how close a fixed orbit can get to that: *can it be eclipse-free for a whole year, at what altitude, and how precisely does it have to be placed?*

This paper answers that question with a small amount of orbital mechanics and one idea — a reference frame that rotates with the average Sun — that makes the geometry easy to see. The companion visualizer at colure.space/kelvinzero draws every picture in this paper live; buttons throughout the text open the matching view.

## 2 Orbits that keep pace with the Sun

### The equatorial bulge and the slow twist of an orbit

If Earth were a perfect sphere — and if the Sun and Moon did not tug on the satellite, which at these altitudes[G1] they do only slightly — an orbit plane would stay fixed in space, pointing at the same distant stars. However, Earth is not a perfect sphere: it spins, and its rotation flings the equator outward so the equatorial radius is about 21 km larger than the polar radius. That extra ring of mass pulls on a satellite a little more when it is near the equator than when it is over the poles. For an orbit that is tilted with respect to the equator, the net effect of that uneven tug, averaged over many revolutions, is not to change the tilt but to slowly *pivot the whole orbit plane around Earth's spin axis* — exactly as a leaning spinning top pivots around the vertical. Astrodynamicists call the bulge's gravitational term $J_2$ and the pivoting *nodal precession* [5].

How fast the plane pivots depends on two things the mission designer chooses: how high the orbit is and how steeply it is inclined to the equator. For a circular orbit of radius $a = R + h$ (Earth's radius $R = 6378$ km plus the altitude $h$) the rate is

$$\dot{\Omega} = -\tfrac{3}{2}\, n\, J_2 \left(\frac{R}{a}\right)^2 \cos i, \qquad n = \sqrt{\mu/a^3}, \tag{1}$$

where $n$ is the satellite's angular speed around Earth, $i$ its inclination[G2], $J_2 = 1.0826 \times 10^{-3}$ the size of the bulge and $\mu$ Earth's gravitational constant. The sign matters: an orbit tilted less than 90° pivots westward, and an orbit tilted *past* 90° — one that travels slightly against Earth's rotation, called retrograde — pivots eastward.

### Matching the Sun's rate

Seen from Earth, the Sun also drifts eastward against the stars, once around per year: $360^\circ/365.24\text{ d} = 0.9856^\circ$ per day. If the orbit plane pivots eastward at that same rate, the angle between the plane and the Sun never changes. The Sun rises over the same ground at the same local time on every pass, which is why Earth-observation satellites love these orbits — and why a plane set to face the Sun keeps facing it. Such an orbit is called **sun-synchronous**. Setting $\dot{\Omega}$ in (1) equal to $\omega_\odot = 0.9856^\circ/\text{day}$ and solving for the inclination gives

$$\cos i_{\text{SSO}} = -\frac{\omega_\odot}{\frac{3}{2}\, n\, J_2 (R/a)^2}. \tag{2}$$

For every altitude there is exactly one inclination that works (Figure 1a, Table 1): about 97.6° at 550 km, 98.6° at 800 km, 99.5° at 1000 km. Altitude and inclination are therefore not independent choices; pick the altitude and the inclination follows. That leaves the designer with two free variables — the altitude $h$ and the *orientation* of the plane around the spin axis — plus the trivial choice of where along the ring the satellite starts.

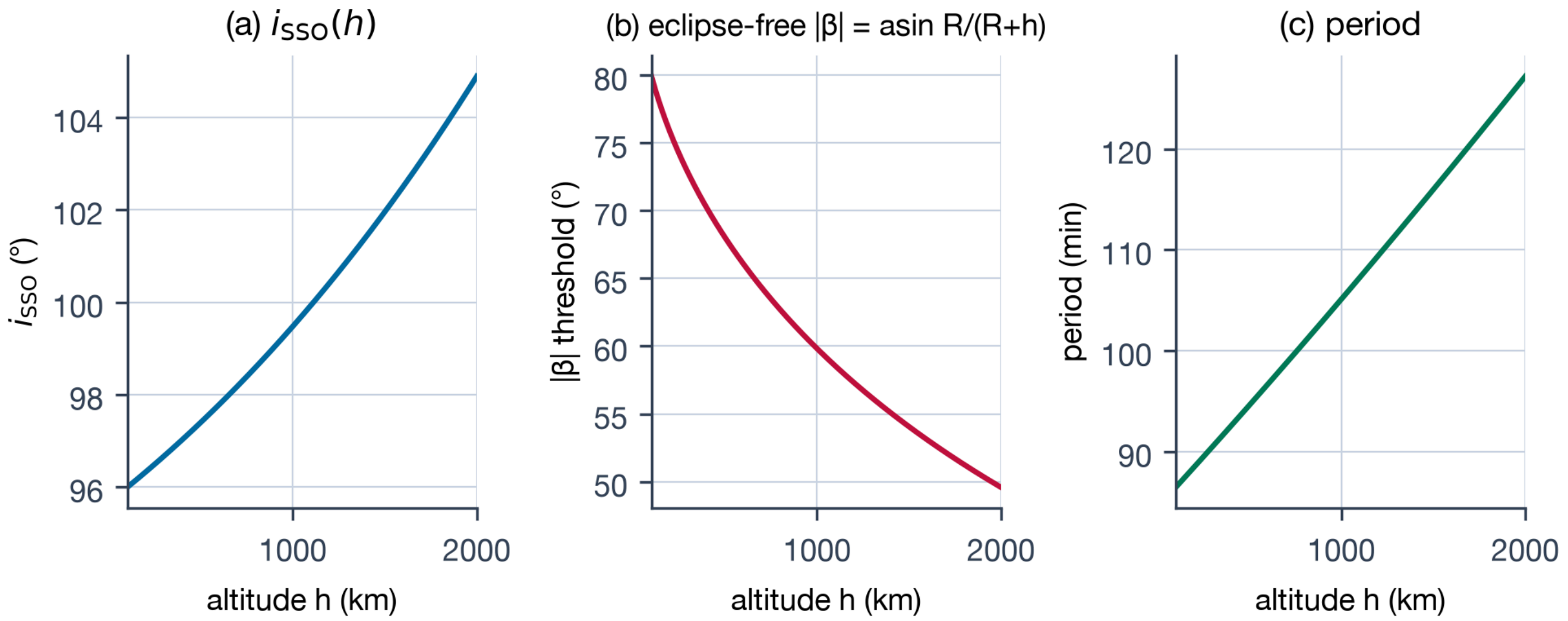

**Figure 1.** Sun-synchronous inclination, the eclipse-free |β| threshold of Section 6 and the orbital period versus altitude for circular orbits.

| h (km) | $i_{SSO}$ (°) | T (min) | \|β\| for no eclipse (°) | α (°) |
|---|---|---|---|---|
| 300 | 96.672 | 90.52 | 72.76 | 107.24 |
| 400 | 97.030 | 92.56 | 70.22 | 109.78 |
| 500 | 97.402 | 94.62 | 68.02 | 111.98 |
| 550 | 97.593 | 95.65 | 67.02 | 112.98 |
| 600 | 97.788 | 96.69 | 66.07 | 113.93 |
| 700 | 98.188 | 98.77 | 64.30 | 115.70 |
| 800 | 98.603 | 100.87 | 62.69 | 117.31 |
| 1000 | 99.479 | 105.12 | 59.82 | 120.18 |
| 1200 | 100.419 | 109.42 | 57.31 | 122.69 |
| 1500 | 101.957 | 115.98 | 54.06 | 125.94 |
| 2000 | 104.890 | 127.20 | 49.58 | 130.42 |

**Table 1.** Sun-synchronous inclination, period, eclipse-free β threshold and region cone half-angle α (Section 7) versus altitude. R = 6378.137 km, $J_2 = 1.08262668\times10^{-3}$, μ = 398600.4418 $km^3/s^2$.

## Orienting the plane: the node offset and local time

The orientation of an orbit plane is described by where it crosses the equator going north, the *ascending node*, and the angle from a reference direction to that node is the right ascension of the ascending node[G6], $\Omega$. For a sun-synchronous[G5] orbit the natural reference is the Sun itself, because the node keeps a fixed angle from it. We call that angle the *node offset* $\Omega_{\text{off}}$ (the slider labelled "RAAN offset from mean Sun[G11]" in the visualizer) and it is equivalent to a clock reading, the local time of the ascending node[G8]:

$$\text{LTAN} = 12{:}00 + \Omega_{\text{off}}/15^\circ. \tag{3}$$

$\Omega_{\text{off}} = 0^\circ$ puts the node under the noon Sun: the plane contains the Sun direction, the satellite flies over the sunlit face and the dark face on every lap and is eclipsed every orbit. $\Omega_{\text{off}} = 90^\circ$ or $270^\circ$ (LTAN 18:00 or 06:00) turns the plane to face the Sun squarely: its normal points at the Sun, sunlight falls perpendicular to the plane, and the satellite flies along the day–night line, always over dawn or dusk. This is the **dawn–dusk orbit**, and it is the obvious candidate for an eclipse-free mission. Figure 2 shows the three quantities together.

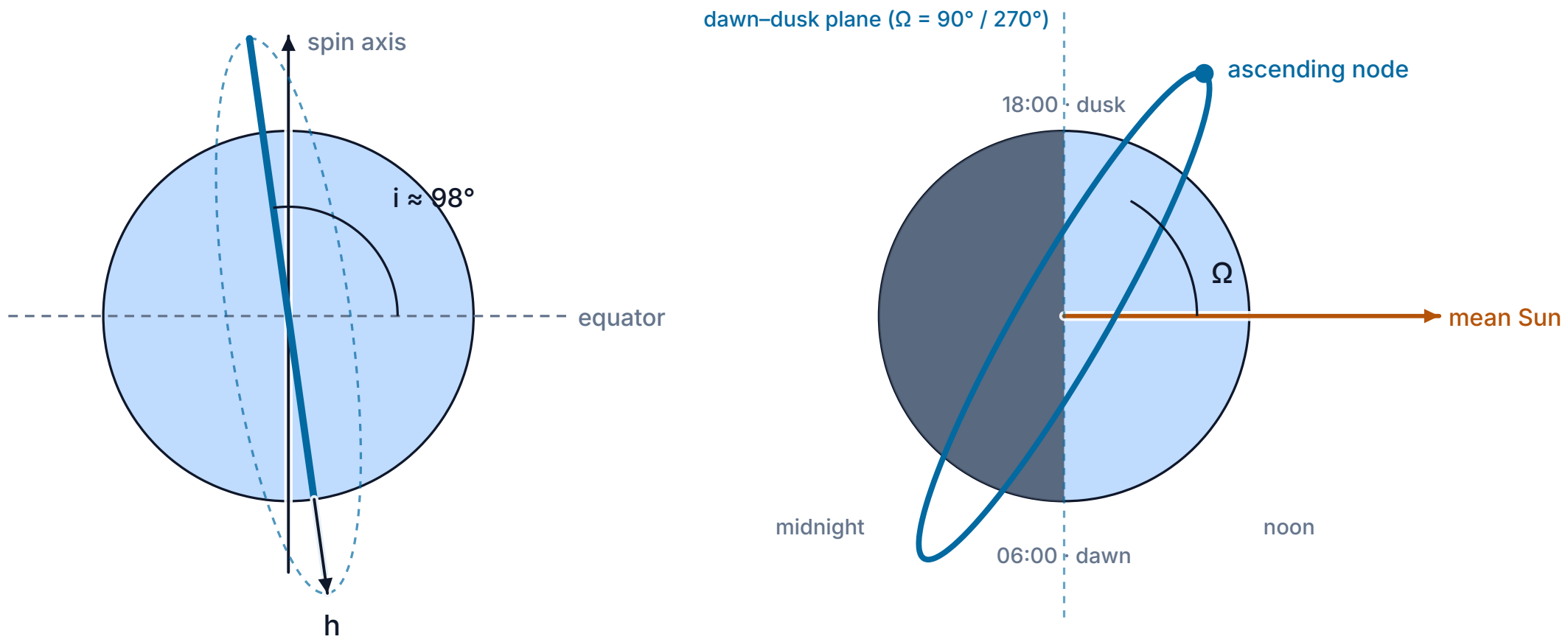

(a) side view: inclination i, set by the altitude h
retrograde: the plane leans past the pole

(b) looking down the spin axis: node offset Ω from the mean Sun
LTAN = 12:00 + Ω/15°

**Figure 2.** The three free variables of a circular sun-synchronous orbit. (a) Side view: the altitude h fixes the inclination i through (2); sun-synchronous orbits are retrograde, leaning a few degrees past the pole. (b) Looking down the spin axis in the mean-Sun frame: the node offset Ω is the angle from the mean-Sun direction to the ascending node, equivalent to the local time of the ascending node; the dawn–dusk plane faces the Sun (its normal is the Sun line).

Interactive: Dawn–dusk plane, 800 km ↗ · Noon–midnight plane, 550 km ↗

## 3 Why dawn–dusk is not quite perfect

Equation (2) matches the orbit to the Sun's *average* motion. The real Sun does not move uniformly across our sky, for two separate reasons.

**Earth's orbit is an ellipse.** Earth is closest to the Sun in early January (perihelion[G28]) and farthest in early July (aphelion). By Kepler's second law it moves fastest when closest, so the Sun's apparent eastward drift is about 7% faster in January than in July. Over a season the Sun runs ahead of, then behind, its average position by up to about 2°.

**Earth's axis is tilted.** The spin axis leans 23.44° from the perpendicular to the orbit plane. The orbit plane of a satellite pivots around that *tilted* axis, but the Sun moves along the *ecliptic*, the plane of Earth's orbit. Projected onto the equator — the only direction the satellite's plane can follow — the Sun's eastward motion speeds up near the solstices and slows near the equinoxes, and the Sun also swings 23.44° north and south of the equator over the year. The north–south swing is the seasons; the east–west speed-up is a second contribution to the Sun running ahead or behind.

Because of these two effects, astronomers long ago defined a convenient fiction called the **mean Sun**: a point that moves eastward along the celestial equator at the perfectly uniform rate of 0.9856° per day. It is the Sun we would have if Earth's orbit were circular and its axis upright. Mean solar time — the time on every clock — is defined by it. A sun-synchronous[G5] orbit is, precisely, an orbit locked to the *mean* Sun: its plane follows the mean Sun[G11] flawlessly and the *real* Sun wanders around that mean position. The size of the wander is the whole story of this paper: about $\pm 4^\circ$ east–west (the famous *equation of time*, $\pm 16$ minutes on a sundial) and $\pm 23.44^\circ$ north–south.

A dawn–dusk[G9] orbit therefore faces the *mean* Sun squarely — its normal points at the mean Sun — and the real Sun shines on it from up to 23° above or below that plane. Whether that is enough to push the satellite into shadow depends on the altitude[G1] — a higher orbit looks over Earth's shoulder more easily — and on how far the node sits from the ideal 06:00 or 18:00 line. The rest of the paper quantifies exactly that.

## 4 Two ways to look at it: the ECI and MSSI frames

To reason about geometry that changes over a year we need a coordinate frame, and the right choice makes the problem almost trivial.

**ECI — Earth-Centred Inertial.** The standard inertial frame of orbital mechanics, used here in its J2000[G19] realisation (the mean equator and equinox of J2000, "EME2000"; the GCRF differs from it by a few hundredths of an arc-second, which is irrelevant at this level). Its origin rides along with Earth around the Sun, but its axes never rotate: $z$ points along Earth's spin axis and $x$ toward a fixed direction among the stars, the vernal equinox[G16] — the direction of the Sun on 20 March. In ECI[G17] a sun-synchronous[G5] orbit plane is seen to pivot once per year, and the Sun circles once per year along the ecliptic[G15].

**MSSI — Mean-Sun Synchronous Inertial.** The frame we introduce for this problem. It is ECI rotated about the spin axis so that its $x$ axis always points at the *mean* Sun. If $L(d)$ is the mean Sun[G11]'s longitude measured from the vernal equinox on day $d$ (days from the J2000 epoch, 2000-01-01 12:00),

$$L(d) = 280.460^\circ + 0.9856474^\circ\, d, \qquad \mathbf{r}_{\mathrm{MSSI}} = R_z(-L)\,\mathbf{r}_{\mathrm{ECI}}. \tag{4}$$

The frame turns once per year — exactly as fast as a sun-synchronous plane precesses. The payoff is that the precession[G4] is *baked into the frame*: in MSSI[G18] a sun-synchronous orbit is one fixed circle around Earth that never moves, the mean Sun is a fixed point on the $+x$ axis, and the only thing that moves over the year is the real Sun, wandering around that point. The yearly eclipse problem becomes a static picture: a fixed ring, and a small loop that the Sun traces beside it. Figure 3 shows both frames carried around the Sun, and Figure 4 and Figure 5 show the Sun-centred view in the visualizer. The visualizer is the best place to see the difference: the two views linked below show the same orbit with the frame planes drawn, once in ECI — press play and the plane visibly pivots as the year passes — and once in MSSI, where it stays put while the Sun wanders.

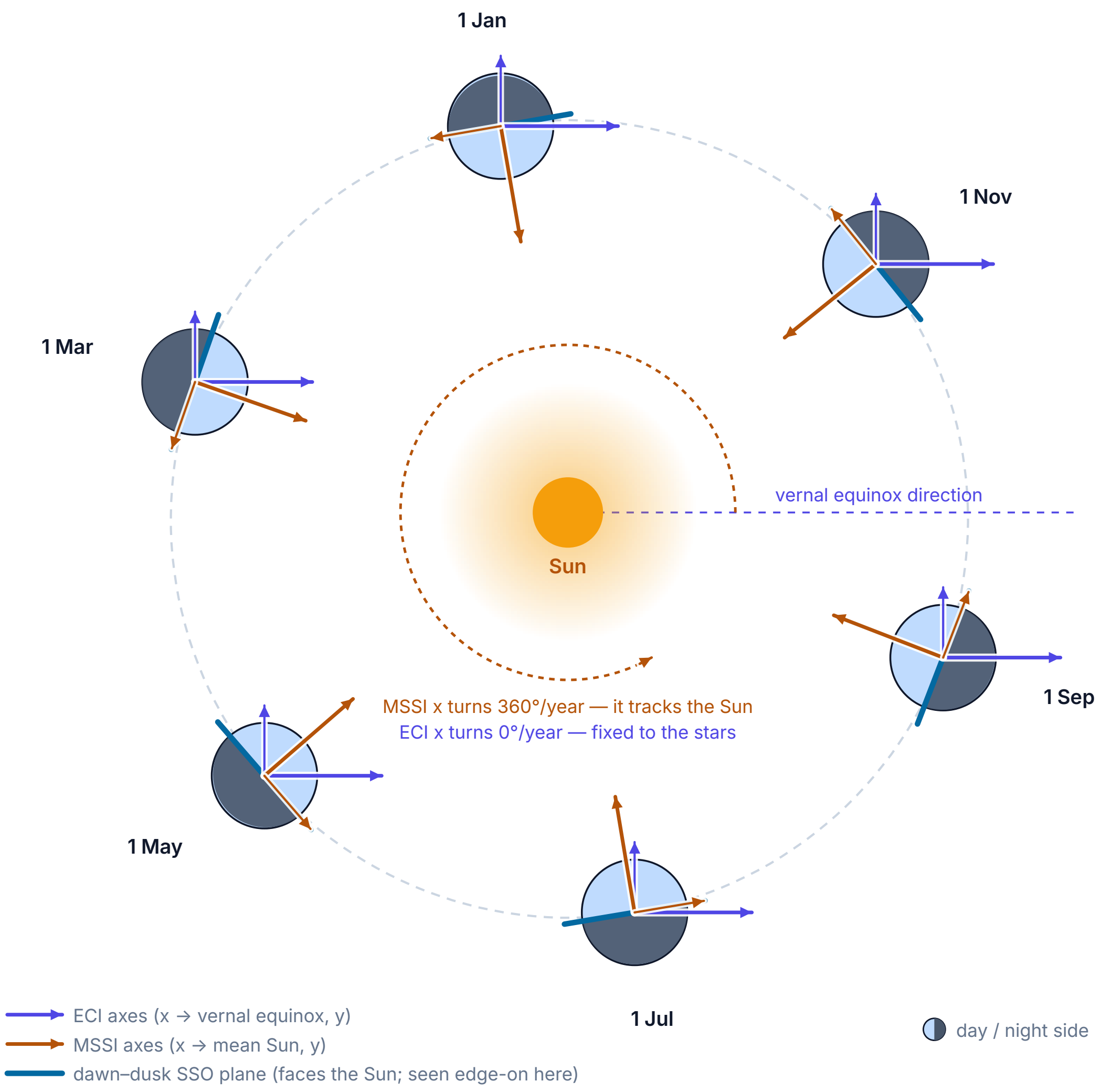

**Figure 3.** The two frames around the year, seen from above the ecliptic (Earth and axes not to scale; orbit from the same ephemeris). On the 1st of every other month the indigo ECI axes point the same way — x toward the vernal equinox — while the amber MSSI x axis has turned to keep pointing at the Sun. A dawn–dusk sun-synchronous plane (sky blue; seen edge-on from above, it faces the Sun) therefore keeps its orientation to the Sun, and the night side of the Earth always lies along −x of MSSI.

Earth's daily rotation is independent of either frame: Greenwich sits at sidereal angle $\theta_G(d) = 280.46^\circ + 360.9856^\circ\, d$ from the equinox, so the globe is turned by $\theta_G - L$ in MSSI. At $d = 0$ (J2000, 12:00 UT) that is $0.0006^\circ$: Greenwich lies under the mean Sun at noon, which is one of the sanity checks used by the end-to-end tests.

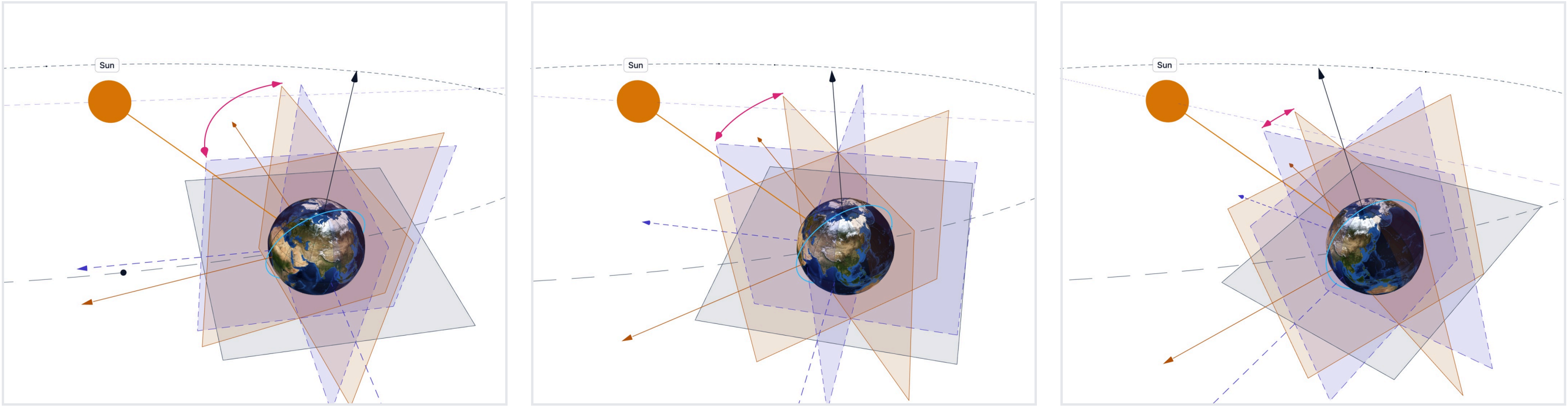

(a) 1 January (b) 1 February (c) 1 March

ECI axes and planes · MSSI axes and planes · equatorial plane (shared) · 800 km dawn–dusk orbit · angle between the ECI and MSSI frames (the mean-Sun longitude L)

**Figure 4.** Sun-centred view, following the Earth around its orbit on the first of three consecutive months: the Sun at the origin, the Earth on its true orbit (dotted; perihelion in early January), an 800 km dawn–dusk plane (Ω = 270°) and the axes and coordinate planes of both frames. From panel to panel the Earth has moved about 30° along its orbit and the amber MSSI x axis has turned by the same 30° to keep pointing at the Sun, while the indigo ECI axes point the same way in all three. The pink arc along the top of the planes marks the angle between the two frames — the mean-Sun longitude L — which closes by about 30° a month toward the March equinox, when the frames coincide; the dawn–dusk ring stays face-on to the Sun throughout. Earth and its frames are enlarged ≈1000×; the orbit is to scale. Open this view in the interactive model ↗

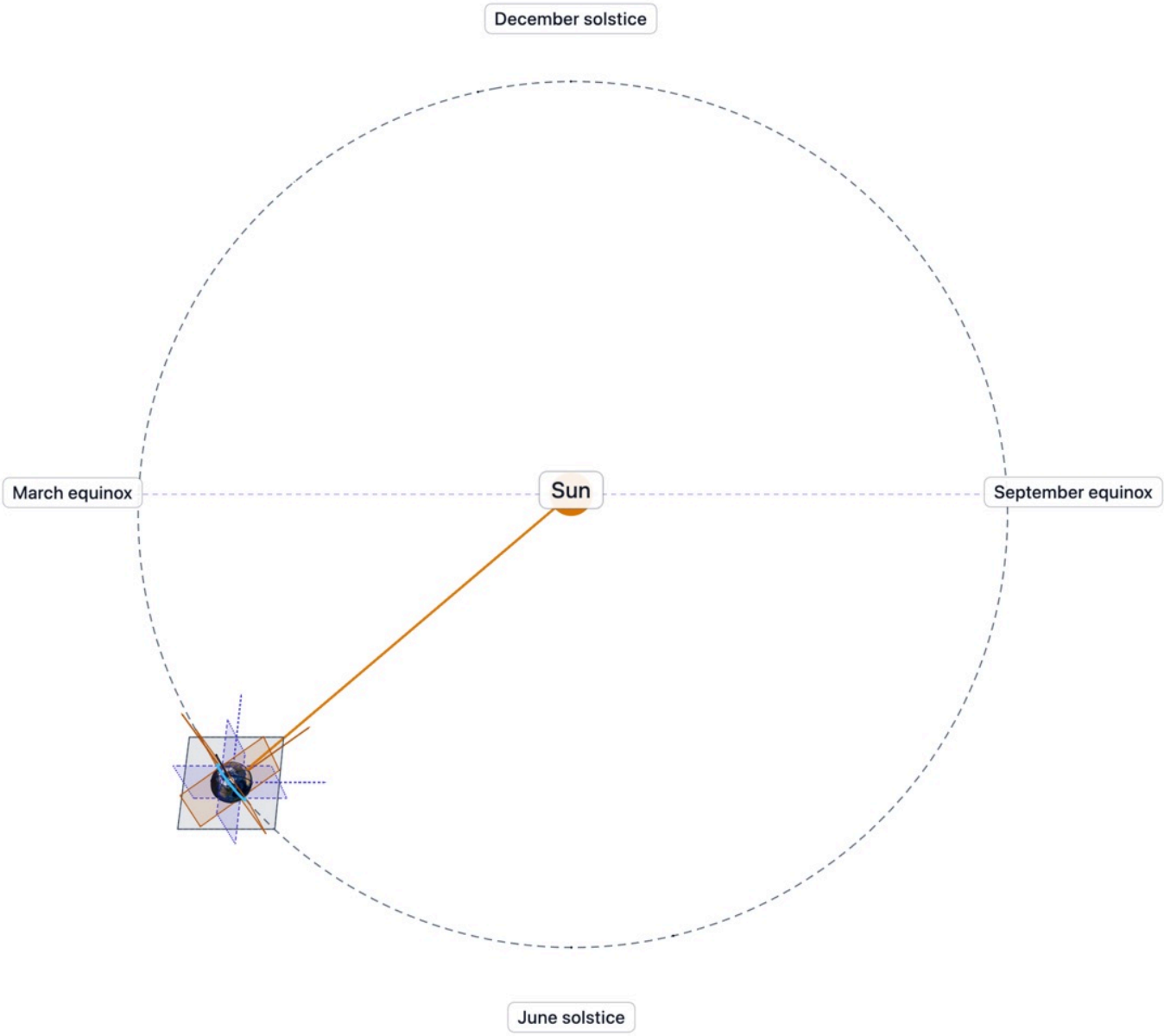

ECI axes and planes · MSSI axes and planes · dawn–dusk orbit · Sun and the Earth–Sun line

**Figure 5.** The full orbit seen from above the ecliptic on 30 April, with the equinox and solstice points marked. The MSSI x axis is the Earth–Sun line and the ECI x axis stays parallel to the fixed vernal-equinox direction (dashed). Open this view in the interactive model ↗

Interactive: Orbit and frame planes in ECI — the plane pivots ↗ · The same in MSSI — the plane stays put ↗ · Sun-centred view (Earth orbiting the Sun) ↗

## 5 The real Sun in MSSI: the analemma

What does the real Sun do in the mean-Sun frame? Its position needs only a short ephemeris — the U.S. Naval Observatory's low-precision solar formulae [1], accurate to about one arc-minute over 1950–2050 — which we use for every Sun-dependent quantity in the model so that the sub-solar point[G20], the Sun vector, the terminator[G21], the eclipse tests and the frame rotation are consistent by construction:

$$g = 357.528^\circ + 0.9856003^\circ d, \qquad \lambda = L + 1.915^\circ \sin g + 0.020^\circ \sin 2g, \qquad \varepsilon = 23.439^\circ, \tag{5}$$

$$\alpha = \operatorname{atan2}(\sin\lambda\cos\varepsilon, \cos\lambda), \qquad \delta = \arcsin(\sin\lambda\sin\varepsilon). \tag{6}$$

Here $g$ is the mean anomaly (where Earth would be on a circular orbit), the $\sin g$ terms are the *equation of centre* — the correction for the elliptical orbit — $\lambda$ is the Sun's true longitude along the ecliptic[G15], $\varepsilon$ the axial tilt, and $(\alpha, \delta)$ are the Sun's right ascension and declination[G13]: its longitude and latitude on the celestial equator. The difference between the real and mean Sun[G11] along the equator is the **equation of time**, $\text{EoT} = L - \alpha$, positive when a sundial runs ahead of the clock [2]. In MSSI[G18] the Sun therefore sits at longitude $\alpha - L = -\text{EoT}$ and latitude $\delta$.

Plot that point over a year and you get a slender figure-8 (Figure 6): the **analemma**. It is the same shape that appears in a photograph of the Sun taken from one spot at the same clock time every week — a picture first captured on film by Dennis di Cicco in 1978–79 — and the same curve printed on old globes and sundials to convert sundial time to clock time. The equation of time[G12] itself has been known since antiquity: Ptolemy tabulated it in the *Almagest* in the 2nd century, and it was the great nuisance of navigation and timekeeping until mechanical clocks replaced the Sun as the standard. The two lobes of the 8 are unequal because the two causes — the elliptical orbit (a once-a-year wave) and the tilt (a twice-a-year wave) — add with different periods; the bigger lobe lies in the southern, northern-winter half of the year, when both effects pull the same way.

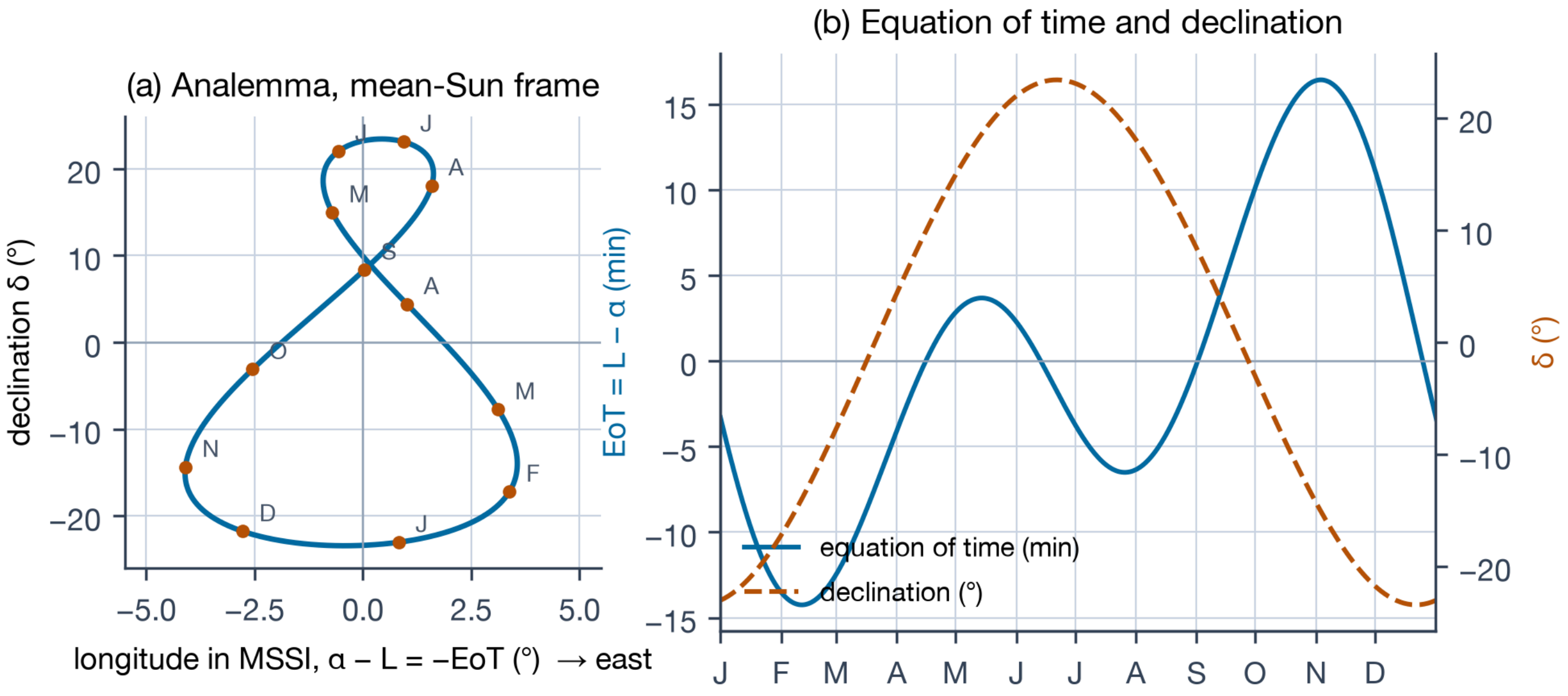

**Figure 6.** (a) The analemma: the locus of the sub-solar point in the mean-Sun frame over one year, with markers on the first of each month. (b) The two components that generate it: the equation of time (the east–west offset, 1° = 4 minutes) and the solar declination (the north–south swing).

There is a simpler way to see why the figure-8 has to appear here. The classic analemma[G14] photograph is made by photographing the Sun from the same spot at the same clock time every week or so. Clock time is mean solar time, so "the same clock time" means "when the mean Sun is in the same place in the sky": the photographer is recording where the real Sun is relative to the mean Sun. That is exactly what the MSSI frame does continuously — it holds the mean Sun still and records the real Sun — so the photographer's figure-8 (Figure 7) and ours are the same curve, drawn from Earth's centre instead of from a roadside. The connection to our problem is now direct. The analemma is the set of directions from which sunlight arrives over the year, drawn in the frame where the orbit is frozen. Its height, $\pm 23.44°$, is the Sun's excursion above and below the plane of a dawn–dusk[G9] orbit; its width, $\pm 4°$, is how far the Sun strays east–west from the direction the plane faces. Figure 8 shows the curve on the globe, with the Sun on it, in the visualizer.

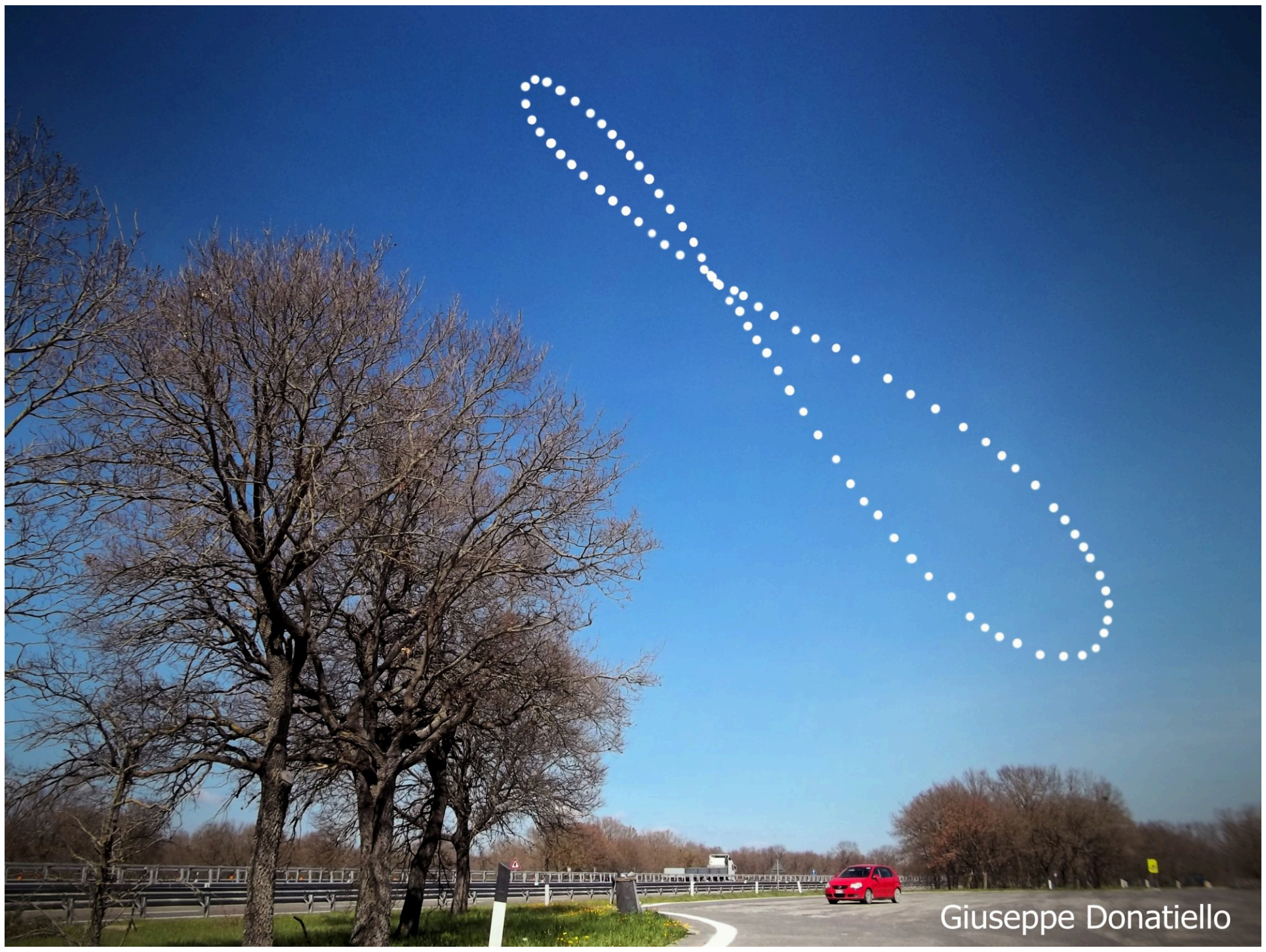

**Figure 7.** The analemma as a photographer sees it: the Sun photographed from the same spot in southern Italy at the same clock time every few days through 2015–16, multiple exposures on one frame. Same clock time means the mean Sun is in the same place each time, so the figure-8 is the real Sun's wander about the mean Sun — the curve of Figure 6, seen from the ground. Photograph by Giuseppe Donatiello, released into the public domain (CC0) [3].

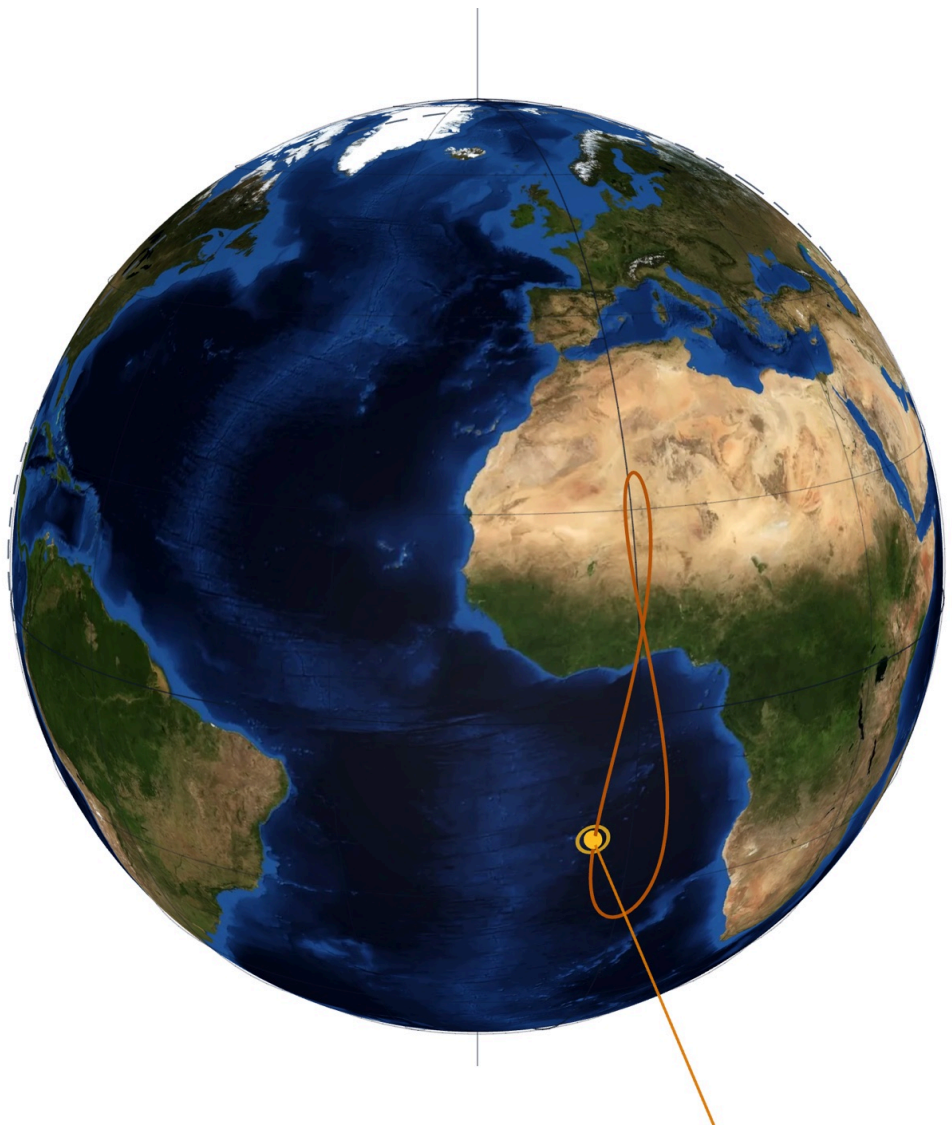

**Figure 8.** The analemma drawn on the sphere in MSSI at day 300 (late October), with the sub-solar point on the curve, the Sun vector and the terminator (dashed). Open this view in the interactive model ↗

Interactive: Analemma at the December solstice ↗ · Analemma at the June solstice ↗

# 6 When does the orbit see the Sun?

With the Sun effectively at infinity, Earth casts a cylindrical shadow[G23] of radius $R$ pointing directly away from the Sun. A point at position $\mathbf{r}$ is in shadow when it is on the night side and within one Earth radius of the shadow's axis:

$$\mathbf{r} \cdot \hat{\mathbf{s}} < 0 \quad \text{and} \quad \|\mathbf{r} \times \hat{\mathbf{s}}\| < R, \tag{7}$$

where $\hat{\mathbf{s}}$ is the unit vector toward the Sun. For a circular orbit the only number that matters is the **β angle**: the elevation of the Sun above the orbit plane, $\sin\beta = \hat{\mathbf{h}} \cdot \hat{\mathbf{s}}$ with $\hat{\mathbf{h}}$ the plane's normal (Figure 9). β lives in the plane spanned by the orbit normal and the Sun direction: turn the picture until you look along the orbit plane with the Sun still in view, and the ring collapses to a line with the Sun β above it, which is how Figure 9b is drawn. When $\beta = 0$ the Sun lies in the plane and the satellite crosses the full width of the shadow; as $\beta$ grows the ring is tilted out of the shadow and the dark arc shrinks, until the whole ring clears the cylinder. Writing $c = R/(R+h)$, the fraction of each revolution spent in shadow is

$$f_E(\beta) = \frac{1}{\pi} \arccos\left(\frac{\sqrt{1-c^2}}{\cos\beta}\right) \quad \text{for } \cos\beta > \sqrt{1-c^2}, \qquad f_E = 0 \text{ otherwise.} \tag{8}$$

The orbit is eclipse-free once $|\beta| \geq \arcsin c$ — 67.0° at 550 km, 62.7° at 800 km, 59.8° at 1000 km (Figure 1b). This is where altitude[G1] enters: a higher satellite sees over Earth's edge at a shallower angle, so it tolerates a Sun that is further from the ideal direction. Figure 10 plots (8); Figure 11 shows the shadow test applied to a noon–midnight[G10] ring in the visualizer.

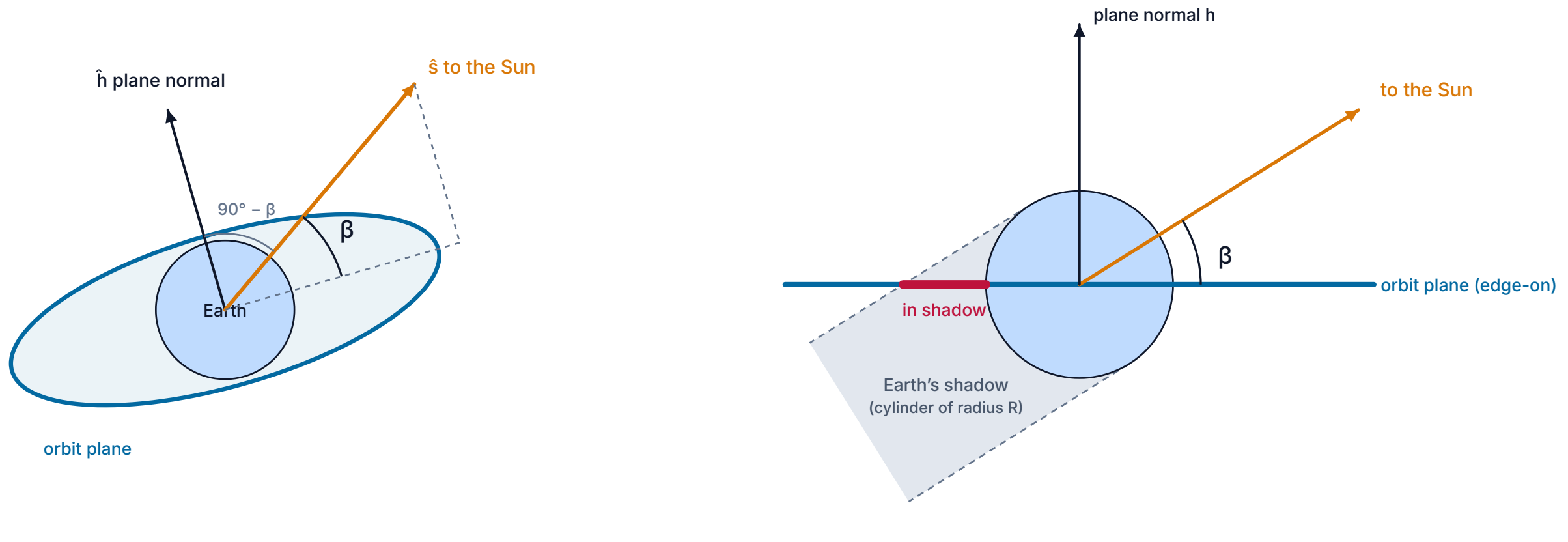

(a) β is the elevation of the Sun above the orbit plane: sin β = ĥ · ŝ

(b) edge-on: the ring clears the shadow once |β| ≥ asin(R/(R+h))

**Figure 9.** The β angle. (a) In perspective: the orbit ring with its normal ĥ; the Sun direction ŝ rises β above the plane, i.e. it makes 90° − β with the normal, measured in the plane that contains ĥ and ŝ. (b) The same seen edge-on — looking along the orbit plane with the Sun in view, so the ring is a line and the Sun sits β above it. Earth's shadow is a cylinder of radius R pointing away from the Sun; the red part of the orbit lies inside it. Tilting the plane (raising |β|) lifts the ring out of the cylinder, and above |β| = asin(R/(R+h)) no part of the ring is ever in shadow.

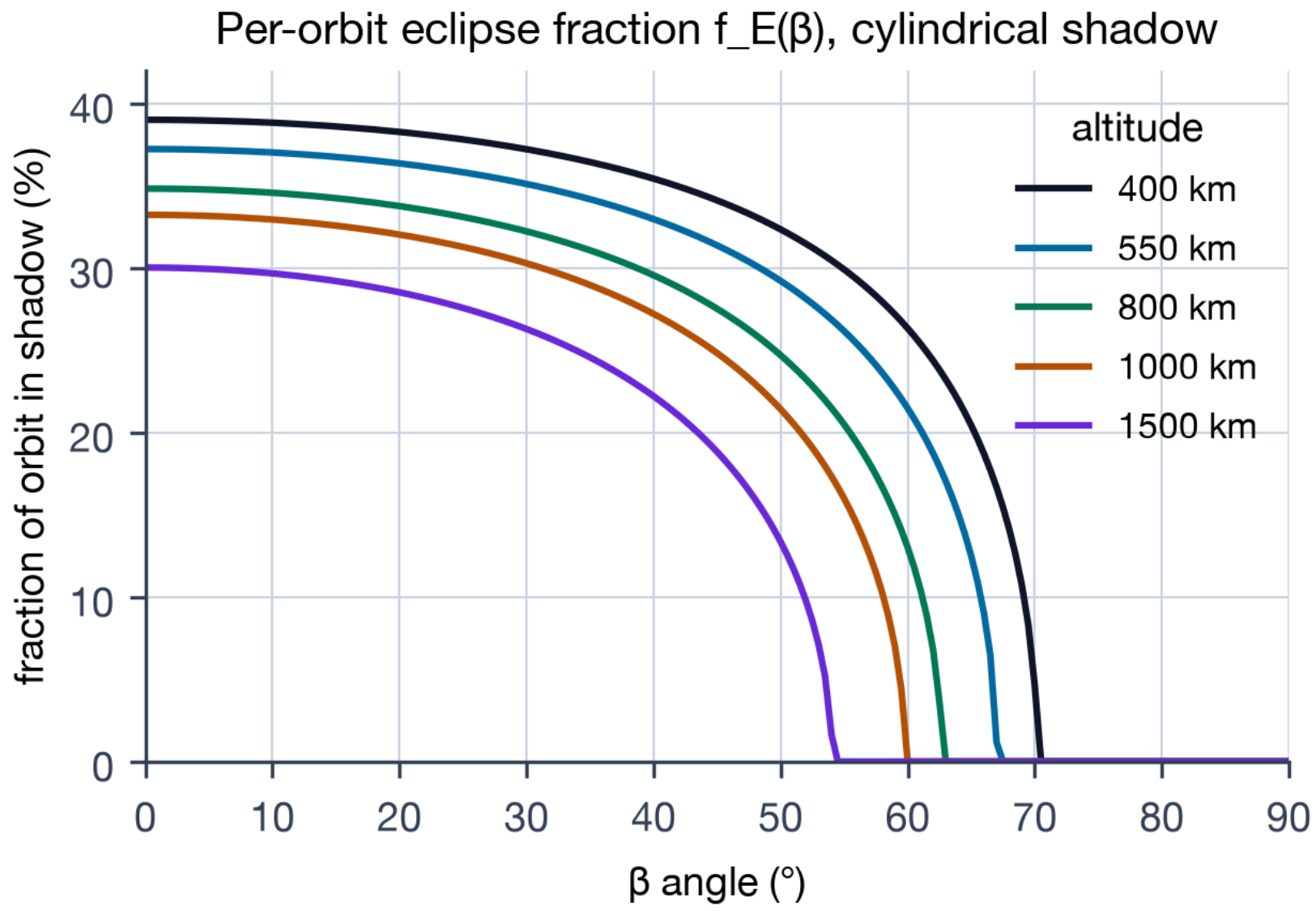

**Figure 10.** Per-orbit eclipse fraction (8) versus β for several altitudes. Higher altitude lowers both the maximum fraction and the β angle at which eclipses vanish.

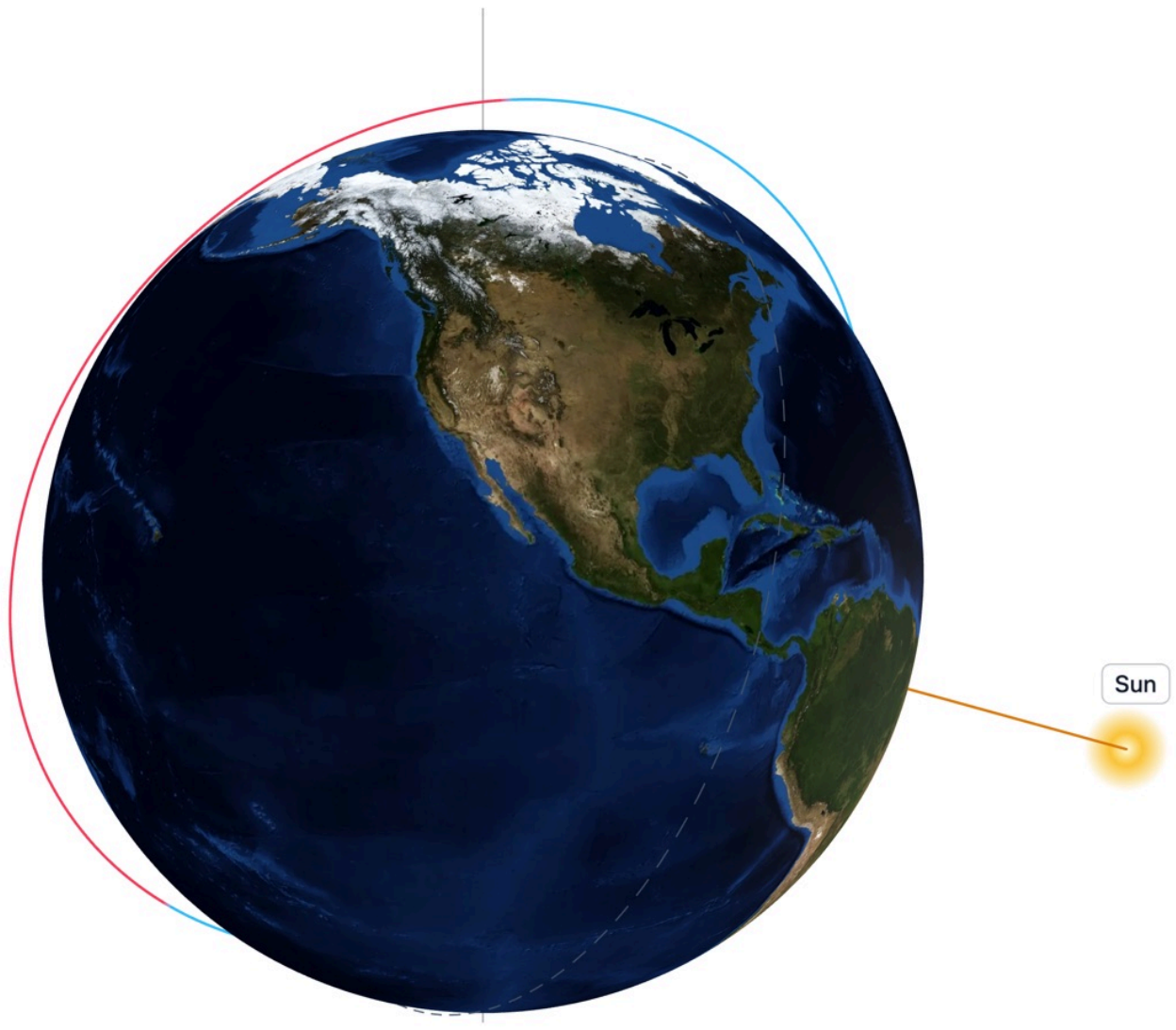

**Figure 11.** Instantaneous shadow test on a 550 km noon–midnight plane (Ω = 0°) at day 0, seen from the night side with the Sun just past the limb: the red arc of the ring satisfies (7); the dashed great circle is the terminator. Open this view in the interactive model ↗

For a dawn–dusk[G9] orbit the normal $\hat{\mathbf{h}}$ points almost at the mean Sun[G11] — not exactly, because a sun-synchronous[G5] orbit is tilted 8° or so past the pole, so the normal sits at latitude $90^\circ - i \approx -8^\circ$. The β angle[G22] is then 90° minus the angle between that normal and the real Sun, which moves around the analemma[G14]. Near one solstice the Sun is 23° on the far side of the normal and β drops to about 58°; near the other it is on the near side and β is nearly 90°. Whether 58° is enough depends on the altitude: at 800 km it is below the 62.7° threshold, so the orbit has a short eclipse season[G26] around that solstice; at 1500 km the threshold is 54° and the orbit never eclipses. The yearly exposure is simply the average of $f_E$ over the days of the year.

Interactive: Eclipse segments from the anti-Sun side ↗

## 7 Maps of eclipse-time regions

The β-angle[G22] argument answers the question one orbit at a time. The mean-Sun frame allows something more general: a map, drawn once per altitude[G1], that shows at a glance which orbits are eclipse-free. Think of the sphere of all directions around Earth, seen from the centre, in MSSI[G18]. A satellite at altitude $h$ lying in direction $\hat{\mathbf{p}}$ is sunlit exactly when the angle between $\hat{\mathbf{p}}$ and the Sun is less than a cone half-angle that depends only on the altitude:

$$\text{sunlit} \iff \angle(\hat{\mathbf{p}}, \hat{\mathbf{s}}) \le \alpha, \qquad \alpha = 90^\circ + \arccos\frac{R}{R+h}. \tag{9}$$

Because the Sun's direction $\hat{\mathbf{s}}(t)$ is a fixed curve in MSSI — the analemma[G14] — the whole year's behaviour of every direction is captured by two numbers, the largest and smallest value of $\hat{\mathbf{p}} \cdot \hat{\mathbf{s}}$ over the year:

$$U = \{\hat{\mathbf{p}} : \max_t \hat{\mathbf{p}} \cdot \hat{\mathbf{s}} \ge \cos\alpha\} \text{ (ever sunlit)}, \qquad I = \{\hat{\mathbf{p}} : \min_t \hat{\mathbf{p}} \cdot \hat{\mathbf{s}} \ge \cos\alpha\} \text{ (always sunlit)}. \tag{10}$$

Each direction is coloured by what a point of the orbit shell there sees over the year: **green** — in sunlight all year (inside $I$); **orange** — in sunlight for part of the year and in shadow for the rest (inside $U$ but outside $I$); **purple** — in shadow all year (outside $U$). The boundaries $\partial U$ and $\partial I$ are contours of those two numbers. An orbit at that altitude is a great circle on this sphere, so the map answers the eclipse question at a glance: *(1) if the whole ring lies in green, the satellite is never eclipsed at any time of the year; (2) if the ring passes through orange but not purple, it has an eclipse season — eclipse-free for part of the year, an eclipse on every revolution for the rest; (3) if the ring crosses purple, it is eclipsed on every revolution all year round.* Note that orange does not mean "a little shadow on every orbit": on a given day a point is either sunlit or not, so an orange crossing means whole seasons with eclipses and whole seasons without. Figure 12 shows the maps for 550 km and 1000 km; Figure 13 drapes one on the globe with a noon–midnight[G10] ring crossing the purple cap. Raising the altitude shrinks the purple and orange regions, which is the map's way of saying what (8) said: higher orbits tolerate more.

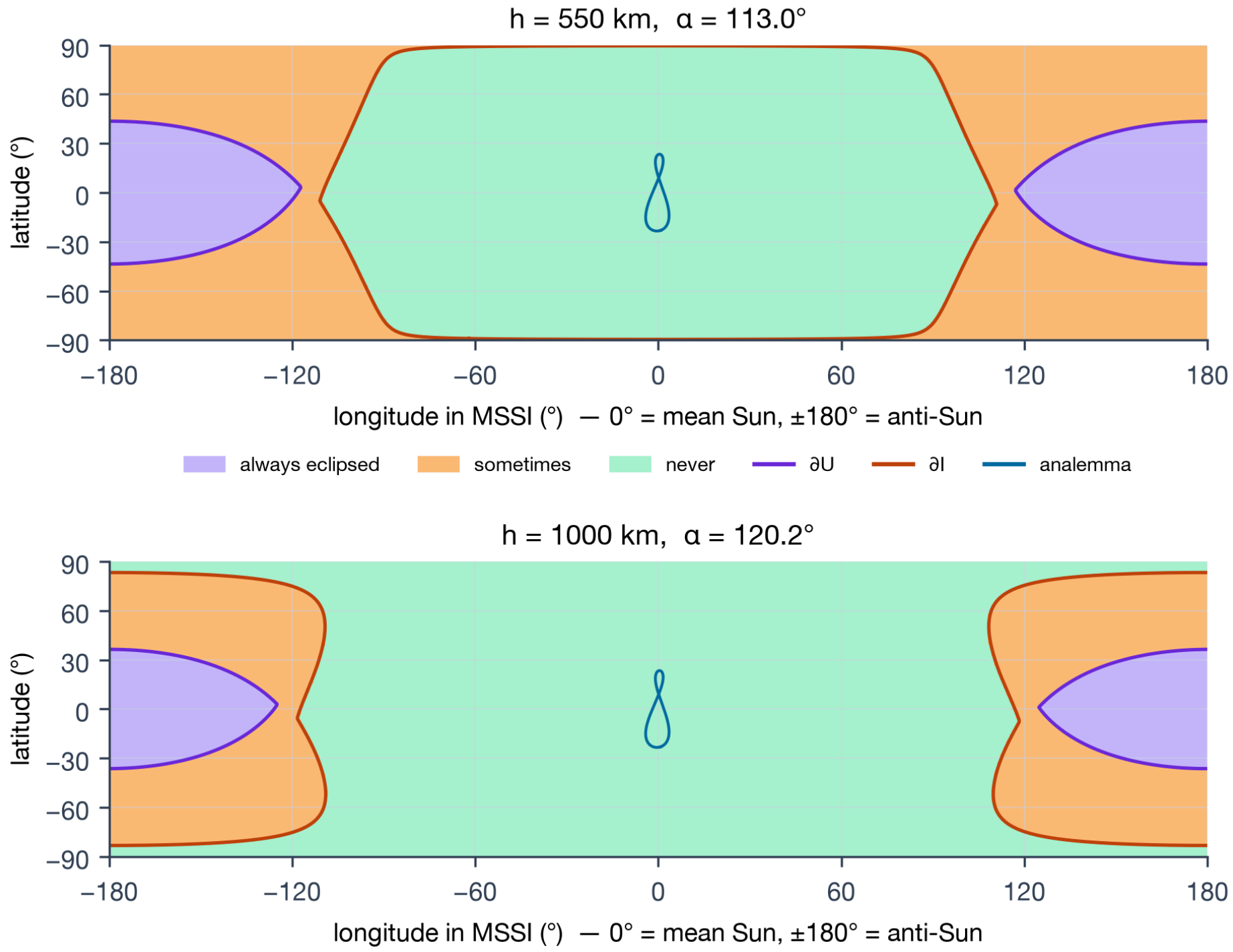

**Figure 12.** Eclipse-time regions in MSSI (equirectangular, 0° = mean Sun) for shells at 550 km and 1000 km. Green: sunlit all year; orange: sunlit for part of the year, in shadow for the rest; purple: in shadow all year. ∂U bounds the purple cap, ∂I bounds the orange band; the analemma is drawn at the centre for scale. An orbit ring entirely in green is never eclipsed; one that enters orange has an eclipse season; one that crosses purple is eclipsed on every revolution all year.

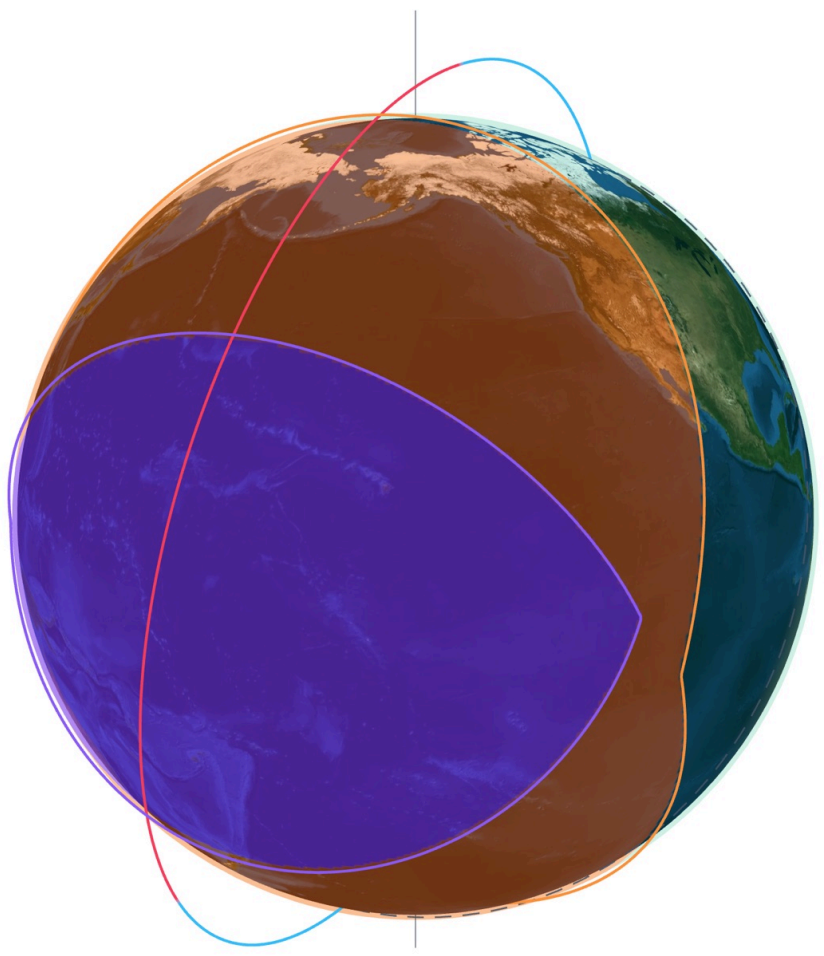

**Figure 13.** The same classification drawn on the sphere for a 1000 km shell, seen from the anti-Sun side. The noon–midnight ring crosses the purple cap, so it is eclipsed on every revolution all year round; the red arc is today's shadow crossing. Open this view in the interactive model ↗

Interactive: Eclipse-time regions ↗ · Regions rotating in ECI ↗

# 8 Results: how much eclipse, and how much margin

## Yearly eclipse versus plane orientation

Figure 14a sweeps the node offset[G7] through a full turn for five altitudes[G1]. The curve has two deep valleys at the dawn–dusk[G9] orientations and two broad hills at noon–midnight[G10], where the plane contains the Sun and the satellite spends 30–37% of the year in shadow. The valleys are not identical. With the orbit normal sitting 8° south of the mean Sun[G11], $\Omega_{\text{off}} = 270°$ (LTAN[G8] 06:00) points the sunward normal north, so its eclipse season[G26] falls around the *December* solstice; because perihelion[G28] is in early January the Sun spends about a week less at southern declinations, and in the closed form 06:00 edges out 18:00: 4.83% versus 5.03% of the year at 550 km, 2.67% versus 2.78% at 800 km. The difference is small — ten hours a year at 800 km — and Section 10 shows that the full force model takes it back: the higher zonal harmonics of the gravity field, which the closed form omits — chiefly $J_3$, which gives the orbit a small forced eccentricity whose orientation relative to the Sun differs between the two lobes — are worth about the same ten hours in the other direction, so the two lobes come out within two hours a year of each other. For eclipse purposes the two dawn–dusk lines are interchangeable; the choice between them can be made on other grounds, such as the lighting of the ground track or ground-station timing. Figure 14b shows the matching β seasons: the 1500 km plane never crosses its threshold.

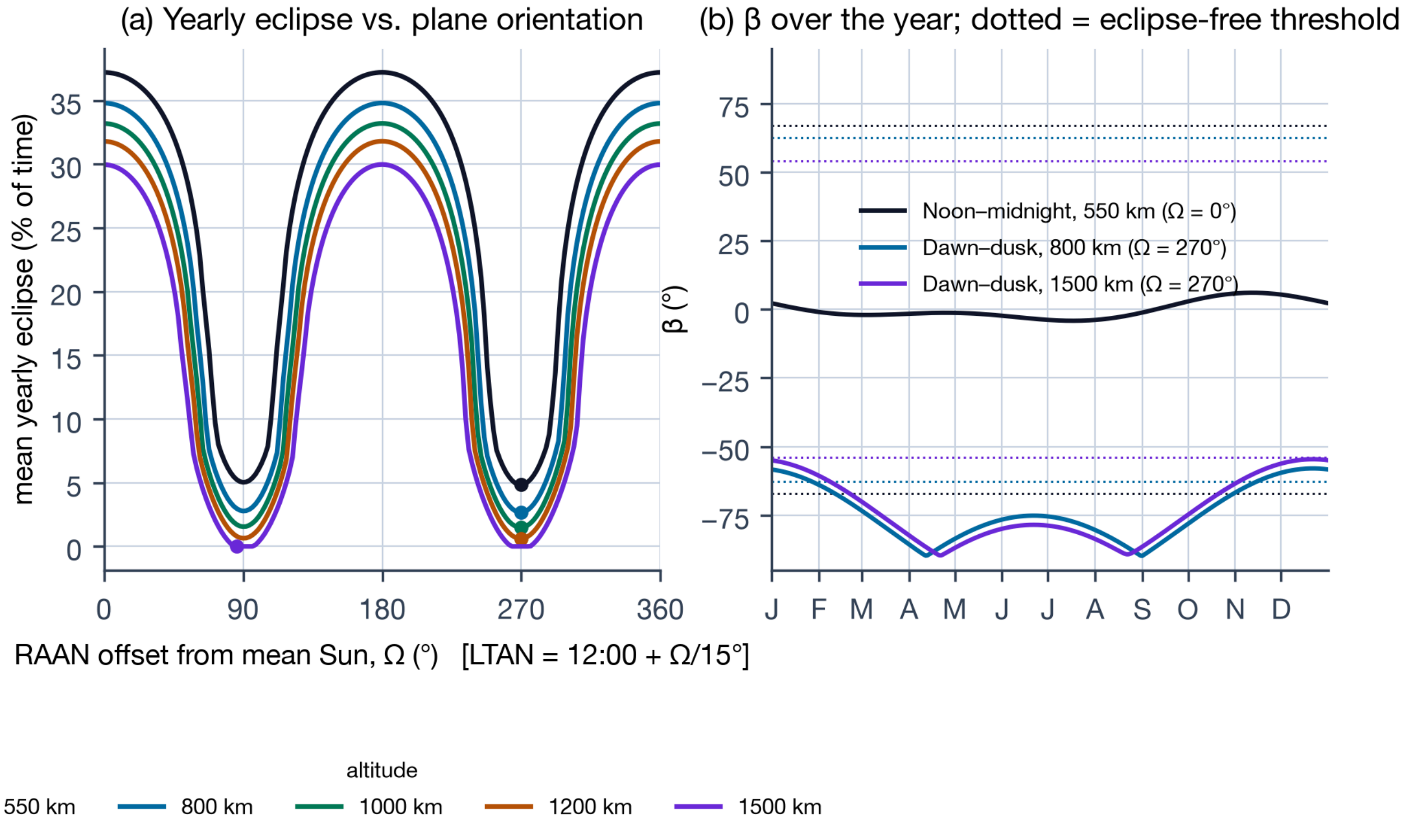

**Figure 14.** (a) Mean yearly eclipse fraction versus node offset Ω for five sun-synchronous altitudes; dots mark each minimum. (b) β over the year for a noon–midnight plane and for dawn–dusk planes at 800 km and 1500 km (Ω = 270°); dotted lines are the eclipse-free thresholds.

## The eclipse season shrinks with altitude

At the best orientation the number of days per year with any eclipse falls from about 97 at 550 km to 74 at 800 km, 56 at 1000 km and 36 at 1200 km (Figure 15). From roughly **1400 km** upward a fixed plane exists whose β never drops below the threshold — a truly eclipse-free sun-synchronous[G5] orbit without any node control. Below that altitude reaching zero would require the plane to follow the Sun's declination[G13] through the year — an "analemma[G14]-following" node schedule, maintained with propulsion — which is outside the scope of a fixed plane.

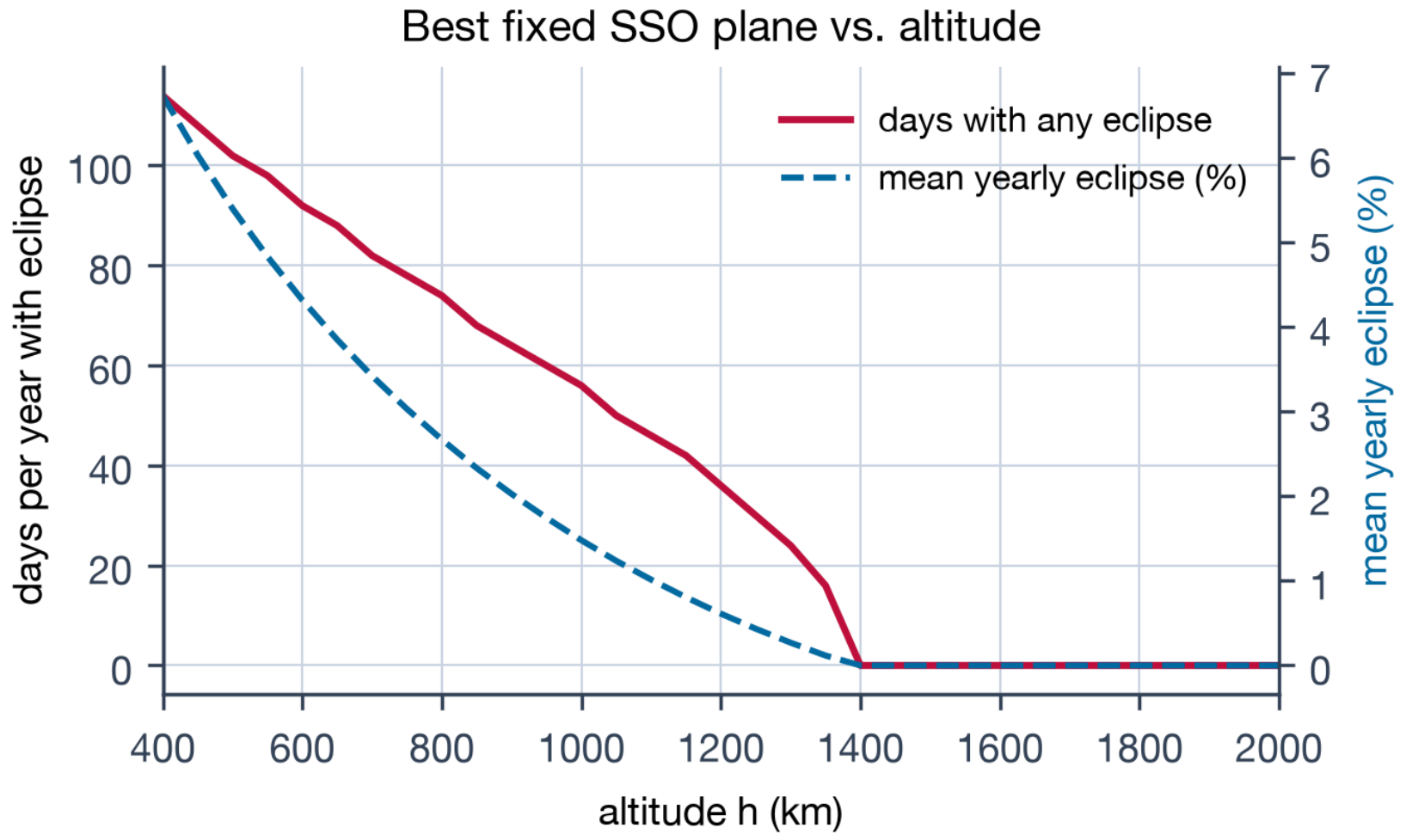

**Figure 15.** Best fixed sun-synchronous plane versus altitude: days per year with any eclipse and mean yearly eclipse fraction at the optimal Ω.

## The full picture, and the margin for error

Figure 16 puts every fixed sun-synchronous plane on one chart: altitude against node offset, coloured by yearly eclipse, with the eclipse-free island outlined in green. Figure 17 reads the margins off that chart: how many degrees the node may sit away from the ideal dawn–dusk line while staying under a given yearly eclipse. The eclipse-free window opens at 1400 km only 3.7° wide (node offsets 267.7°–271.4°), grows to 12.4° at 1500 km (263.4°–275.8°), 16.8° at 1600 km and 25.6° at 2000 km (256.8°–282.4°); it sits almost symmetrically about the 270° line, a few tenths of a degree low. If a mission can accept 1% of the year in shadow (about 90 hours) a window exists from about 1110 km, 10° wide at 1150 km and 14° at 1200 km; at 2% it opens near 900 km, and at 5% — still only 440 hours a year — a 600 km orbit has a 16°-wide window (about ±8° around the dawn–dusk line) and an 800 km orbit 32° (about ±16°). In local time, 15° is one hour of LTAN. Figure 18 shows the same map as a relief. Table 2 collects these design points.

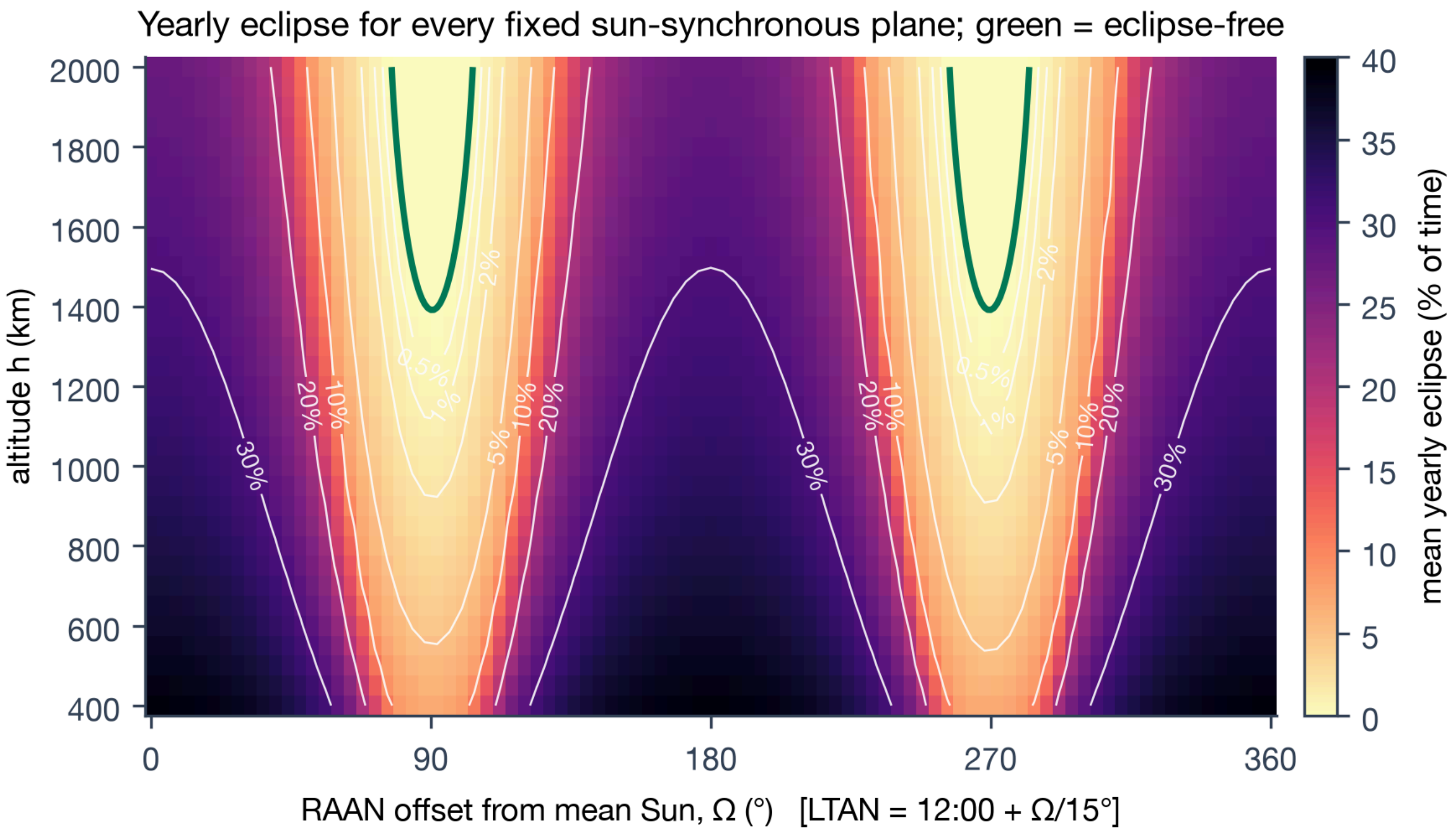

**Figure 16.** Mean yearly eclipse for every fixed sun-synchronous plane, as a function of altitude and node offset. White contours mark 0.5–30%; the green outline encloses the eclipse-free planes, which exist only above about 1400 km.

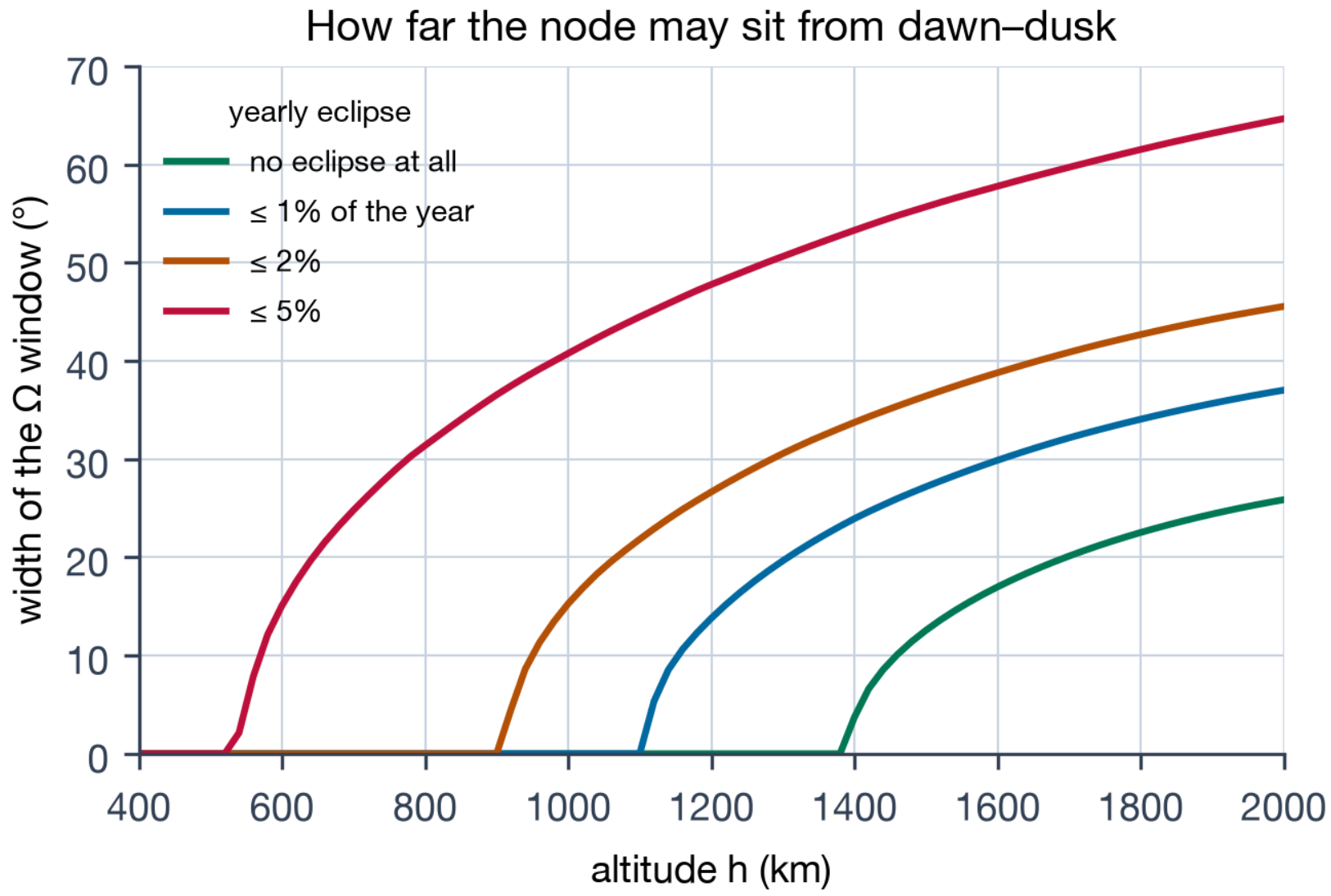

**Figure 17.** Margin for error: the full width of the range of node offsets, around each dawn–dusk line, within which the yearly eclipse stays below a given level — halve it for a ± tolerance about the centre of the range. 15° of node offset is one hour of local time.

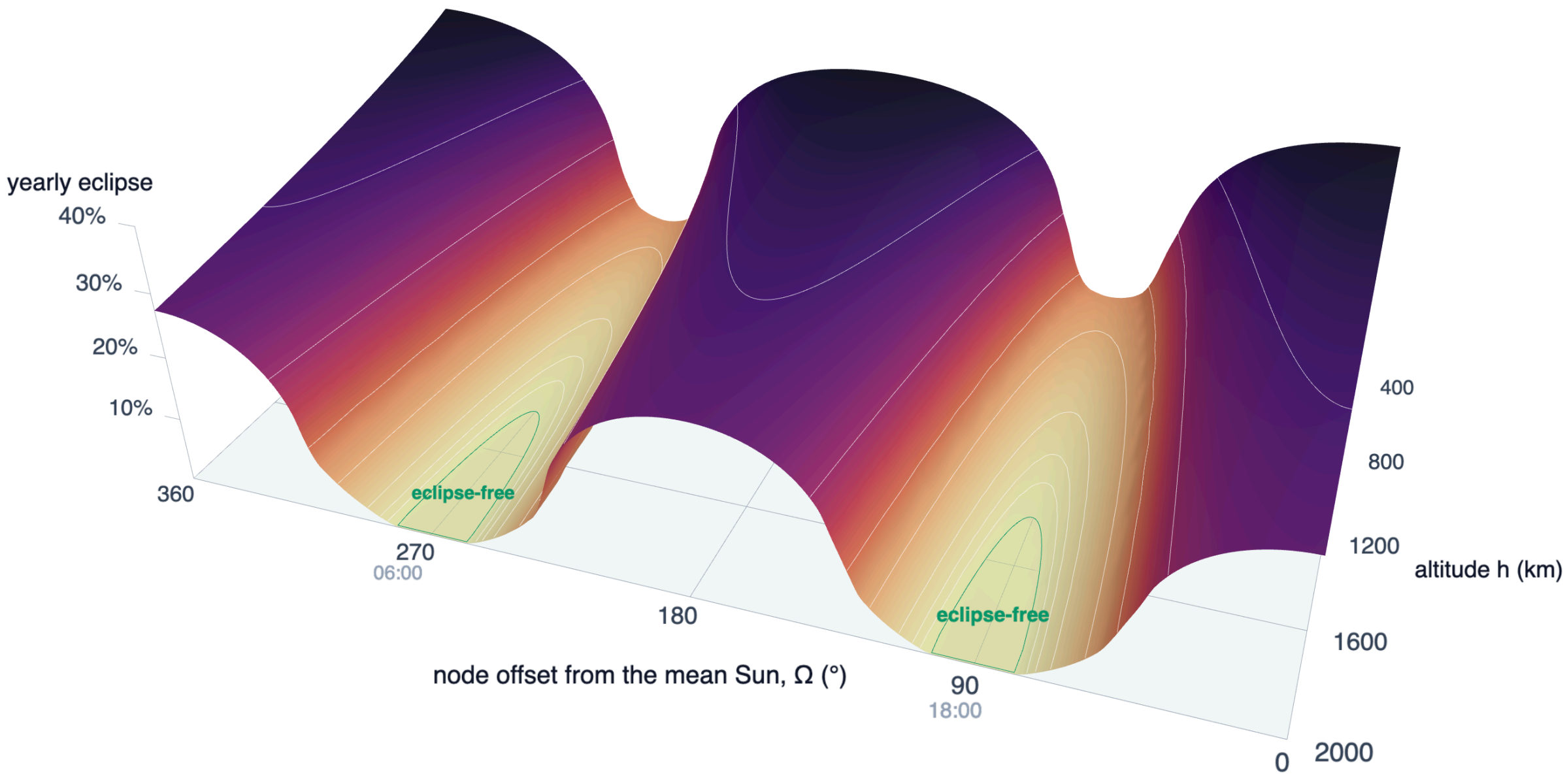

**Figure 18.** The same map as a relief, seen from the high-altitude edge: height and colour are the yearly eclipse, the white lines are the contours of the map above and the green outline on the floor is the eclipse-free island. The dawn–dusk valleys at Ω = 90° and 270° deepen with altitude until, above about 1400 km, they reach the floor. In the web version the relief can be turned by dragging and hovered for values.

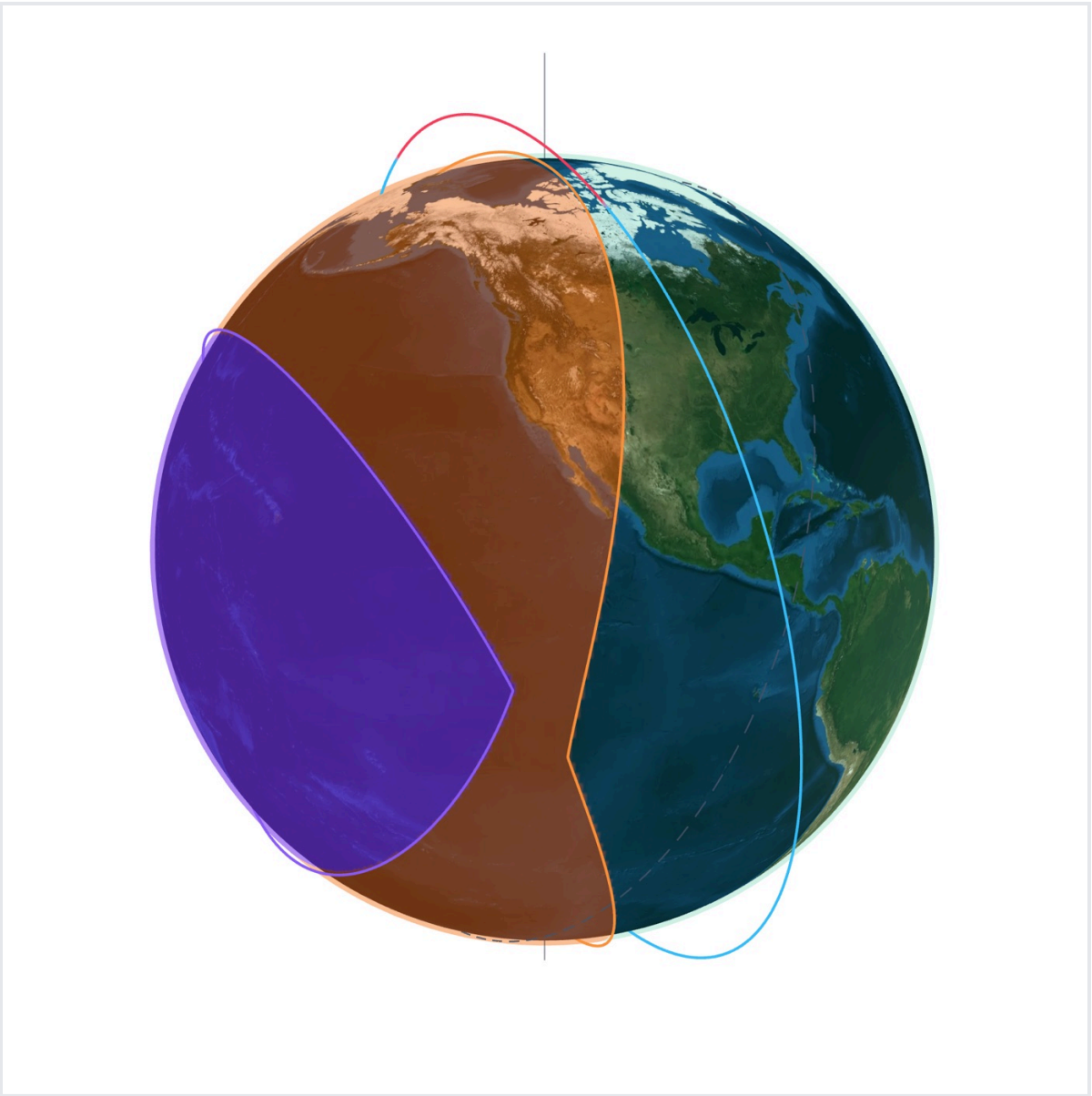

(a) 800 km, Ω = 270°, December solstice

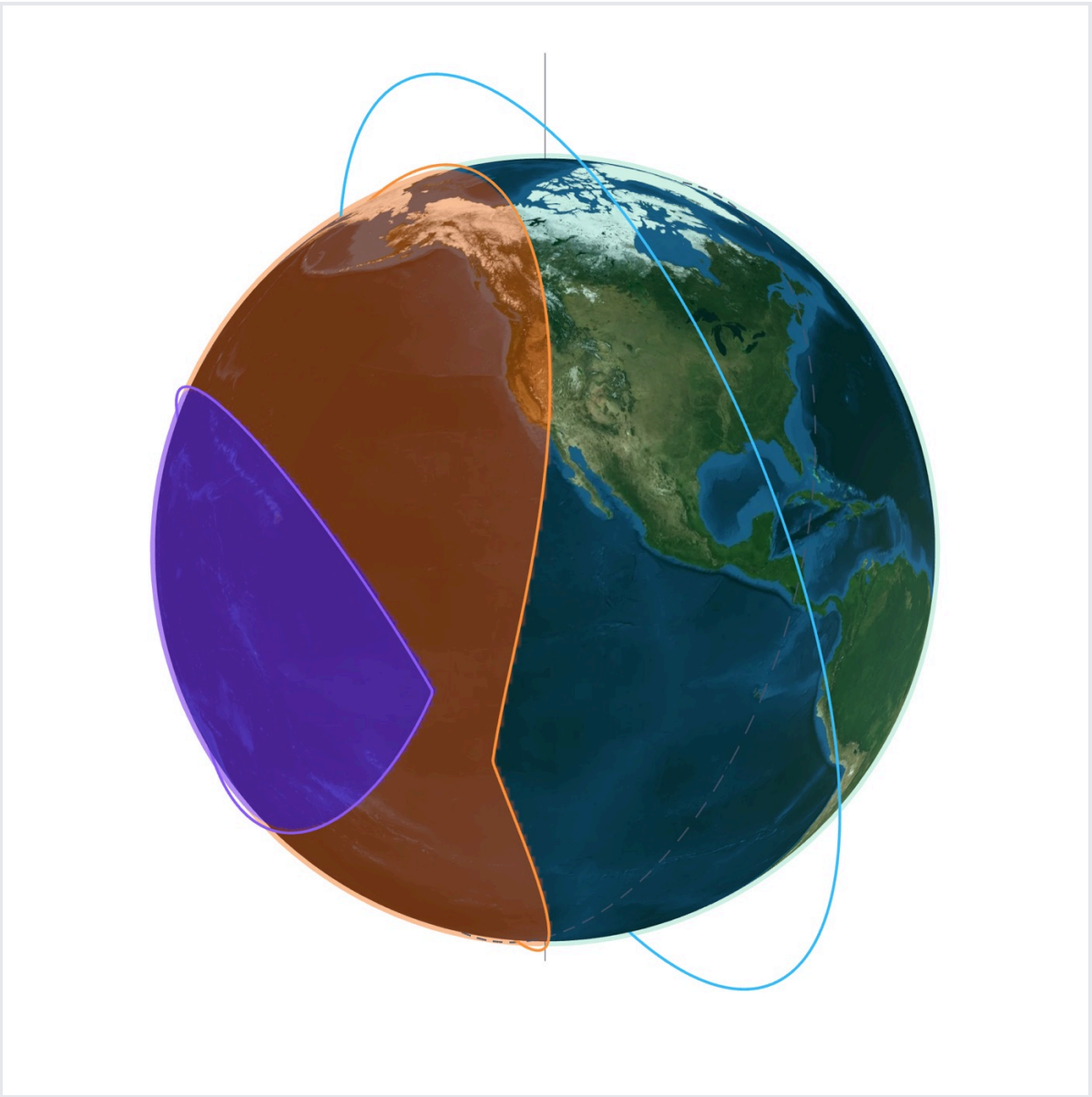

(b) 1500 km, Ω = 270°, December solstice

**Figure 19.** Dawn–dusk planes at their worst time of year, seen from the dawn side with the eclipse-time regions of each shell. (a) At 800 km the ring passes through the orange band, so it has an eclipse season: eclipse-free for most of the year, but on this day (near the December solstice) a short red arc of shadow on every revolution. (b) At 1500 km the ring stays entirely in green: no eclipse at any time of the year. Open this view in the interactive model ↗

Figure 19 shows two dawn–dusk planes on their worst day of the year against the eclipse-time regions[G27] of their shells. In the units a power or thermal engineer works in, the longest eclipse of the year at the best orientation is 21.7 minutes per 95.6-minute orbit at 550 km, 16.9 minutes at 800 km, 13.0 at 1000 km and 8.7 at 1200 km — against about 35 minutes every orbit for a noon–midnight plane — and the yearly totals are 424, 234, 130 and 54 hours. From 1400 km the longest pass is zero. Table 2 lists these numbers with the inclination[G2], the β margin and the node window for each altitude.

| h (km) | i (°) | T (min) | min \|β\| (°) | threshold (°) | eclipse (h/yr) | days | longest pass (min) | eclipse-free Ω (°) | width (°) | ≤ 1 % window (°) |
|---|---|---|---|---|---|---|---|---|---|---|
| 550 | 97.59 | 95.6 | 59.0 | 67.0 | 424 | 97 | 21.7 | — | — | — |
| 800 | 98.60 | 100.9 | 58.0 | 62.7 | 234 | 74 | 16.9 | — | — | — |
| 1000 | 99.48 | 105.1 | 57.1 | 59.8 | 130 | 56 | 13.0 | — | — | — |
| 1200 | 100.42 | 109.4 | 56.1 | 57.3 | 54 | 36 | 8.7 | — | — | 14.0 |
| 1400 | 101.43 | 113.8 | 55.1 | 55.1 | 0 | 0 | 0 | 267.7 – 271.4 | 3.7 | 24.0 |
| 1500 | 101.96 | 116.0 | 54.6 | 54.1 | 0 | 0 | 0 | 263.4 – 275.8 | 12.4 | 27.2 |
| 1600 | 102.51 | 118.2 | 54.1 | 53.1 | 0 | 0 | 0 | 261.2 – 278.0 | 16.8 | 29.8 |
| 1800 | 103.66 | 122.7 | 52.9 | 51.3 | 0 | 0 | 0 | 258.4 – 280.7 | 22.3 | 34.0 |
| 2000 | 104.89 | 127.2 | 51.7 | 49.6 | 0 | 0 | 0 | 256.8 – 282.4 | 25.6 | 36.9 |

**Table 2.** Design points on the 06:00 dawn–dusk line (Ω = 270°). min|β| is the closest the Sun comes to the orbit plane over the year and the threshold is asin(R/(R+h)); the yearly eclipse, days with eclipse and longest pass are from the closed form; the last three columns give the range of node offsets that is eclipse-free, its width, and the width of the range that stays under 1% of the year.

Interactive: Kelvin Zero plane, 1500 km dawn–dusk ↗ · 800 km dawn–dusk in its eclipse season ↗

## 9 A simple model you can use

Everything above reduces to a short recipe that needs no simulation. Given an altitude[G1] $h$ and a node offset[G7] $\Omega_{\text{off}}$:

1. Compute the sun-synchronous[G5] inclination[G2] from (2) and the plane normal $\hat{\mathbf{h}} = (\sin i \sin \Omega,\ -\sin i \cos \Omega,\ \cos i)$ with $\Omega = L(d) + \Omega_{\text{off}}$ in ECI[G17], or simply $\Omega = \Omega_{\text{off}}$ in MSSI[G18].
2. For each day of the year, get the Sun direction from (5)–(6) — in MSSI, $(\cos\delta\cos(\alpha - L),\ \cos\delta\sin(\alpha - L),\ \sin\delta)$ — and the β angle[G22] $\sin\beta = \hat{\mathbf{h}} \cdot \hat{\mathbf{s}}$.
3. Evaluate the per-orbit shadow fraction $f_E(\beta)$ from (8) with $c = R/(R+h)$.
4. Average $f_E$ over the year for the yearly eclipse fraction[G25]; count the days with $f_E > 0$ for the length of the eclipse season[G26]; the orbit is eclipse-free iff $\min_d |\beta(d)| \geq \arcsin c$.

The visualizer's sidebar performs this calculation live for the active orbit — period, LTAN[G8], the current β and shadow fraction, the eclipse-free threshold, the β range and sun exposure over the year, and the number of days with eclipse — and draws β over the year with the eclipse band shaded. The model is de-liberately simple: a circular orbit, J2[G3] secular precession[G4] only, a cylindrical shadow[G23] and the Sun at infinity. Section 11 lists what that leaves out and Section 10 measures it: against a year of numerical propagation with the full gravity field, the Sun and Moon and a conical shadow, the closed form is high by 8–23 hours a year in umbra[G24] time (0.1–0.26% of the year) and within about 8 hours once the penumbra is counted, and neither the season dates nor the eclipse-free threshold move.

## 10 Verification

**Two independent implementations.** The model behind this paper is implemented twice: once in JavaScript, which drives the visualizer and every figure here, and once in Python, written separately from the same equations. The Python implementation produces reference values — the solar ephemeris, the sun-synchronous[G5] inclination[G2] and precession[G4] rate, Kepler-solved and series analemmas, brute-force cylindrical-shadow sweeps over sampled orbits, and the eclipse-region extrema — and an automated test suite compares the JavaScript model against them on every change. The two agree on the solar ephemeris and the sun-synchronous inclination to $10^{-9}$ degrees, on yearly eclipse fraction[G25]s to within 0.0002% of the year — about a minute — (closed form (8) against brute-force sampling of the cylindrical shadow[G23] at 7200 points per orbit on 3660 days), on the eclipse-region extrema to $2 \times 10^{-3}$, and on the analemma[G14] extents to 0.06°. Orientation conventions are tested explicitly: the Sun is west of the mean Sun[G11] in November, an exact sun-synchronous plane is frozen in MSSI[G18], and $\Omega_{\text{off}} = 90°$ gives LTAN[G8] 18:00.

**Scene against model.** An end-to-end harness drives the visualizer's interface and reads world-space positions back from the rendered scene: the sub-solar marker lies on the analemma on every sampled day in both frames; the ring's vertices move by less than $10^{-3}$ Earth radii over a year in MSSI and its node advances 89.70° in 91 days in ECI[G17] (expected 89.69°); the satellite stays at radius $R + h$; the globe's rotation is 0.0006° at $d = 0$ and 180.0° at $d = 0.5$, i.e. Greenwich is under the mean Sun at J2000[G19] noon.

**Against the literature.** The USNO low-precision solar formulae are compared with the widely used NOAA/Spencer compact series [4] over a full year: they differ by at most 0.9 minutes in the equation of time[G12] and 0.5° in declination[G13], both well below anything that affects the conclusions. The closed-form eclipse fraction (8) is the standard result for a circular orbit in a cylindrical shadow [6]; it is worth noting that a tempting simplification, $\arccos(c/\cos\beta)/\pi$ with threshold $\cos\beta > c$, is wrong — it predicts no eclipse beyond $|\beta| \approx 23°$ at 550 km, where the geometric threshold is 67° — and (8) is confirmed against the brute-force sweeps to within 0.0002% of the year.

### Against a full force model

The closed form is only as good as its three simplifications — J2[G3] secular precession, a cylindrical shadow, a one-arc-minute solar ephemeris — so we propagated nine orbits numerically through a full year with none of them: Earth's zonal gravity field through $J_6$ (EGM96), the Sun and Moon as point masses with positions from the JPL DE440 ephemeris, DOP853 integration at a relative tolerance of $10^{-10}$, and every 10 s a conical umbra[G24]/penumbra test with the real Sun–Earth distance and solar disc (pass edges refined to a millisecond). Each case is run twice: with the inclination of equation (2), which measures how far the J2-only sun-synchronous condition is from the truth, and with that inclination corrected once so that the node is stationary relative to the mean Sun under the full force model — what a mission would fly. The whole procedure is `kelvinzero hifi` in the open package and takes a few minutes per case. Figure 20 overlays the daily eclipse minutes of the two models for the 800 km and 1200 km dawn–dusk[G9] planes; Table 3 collects the year totals.

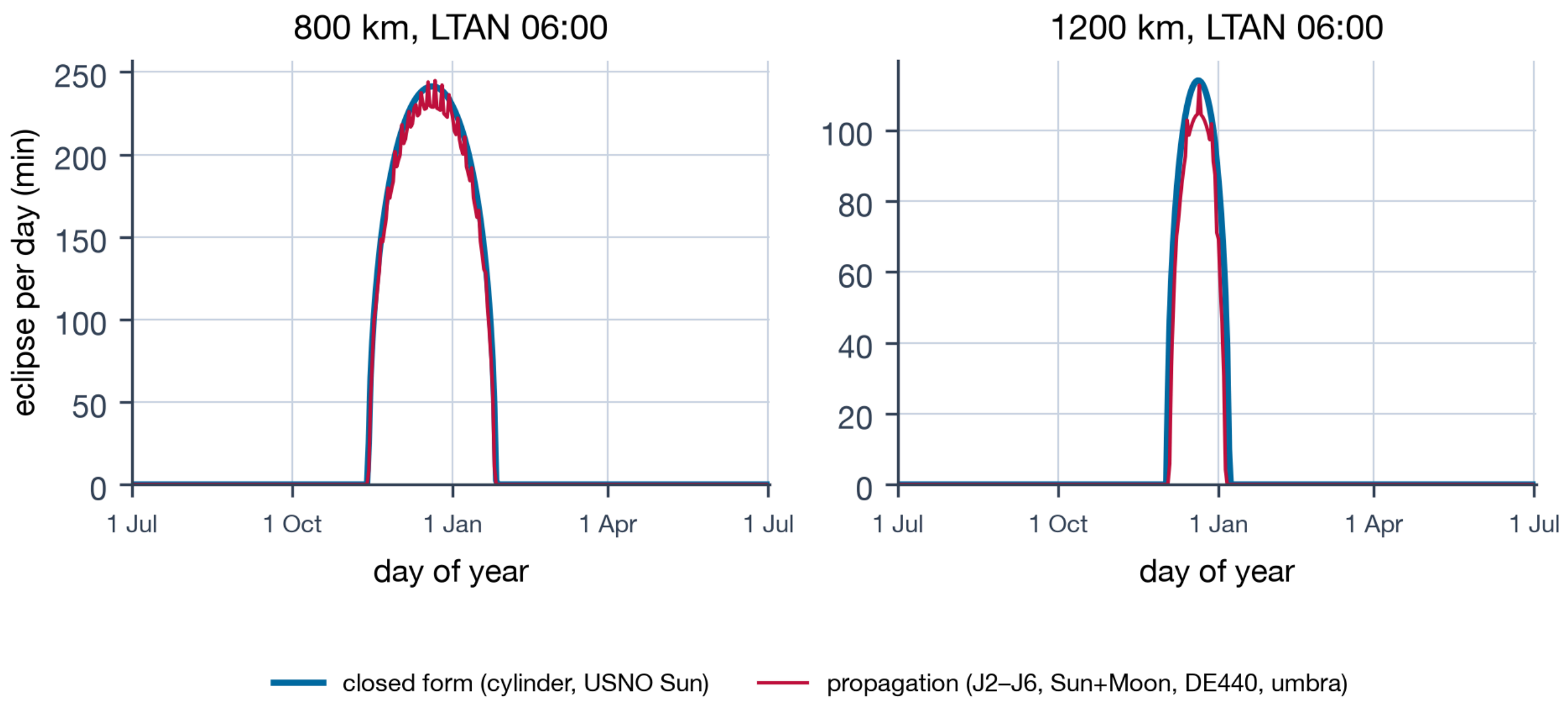

**Figure 20.** Eclipse minutes per day over a year running from July to June (so the December season is whole), closed form against the full-force-model propagation (umbra), for the 800 km dawn–dusk plane and the 1200 km plane, the highest with an eclipse season. The season dates coincide; the propagation runs a few seconds per pass shorter because the umbra is a cone, not a cylinder, and its day-to-day ripple is the 14-or-15 passes that fall in a given day.

| h (km) | LTAN | closed form (h/yr) | days | longest (min) | hi-fi umbra (h/yr) | days | longest (min) | penumbra (h/yr) | drift (°/yr) | Δi (°) |
|---|---|---|---|---|---|---|---|---|---|---|
| 550 | 12:00 | 3261.8 | 365 | 35.6 | 3245.2 | 365 | 35.5 | 26.0 | -0.62 | +0.0132 |
| 550 | 06:00 | 423.8 | 97 | 21.7 | 411.2 | 97 | 21.4 | 31.4 | -0.45 | +0.0095 |
| 800 | 06:00 | 233.9 | 74 | 16.9 | 222.5 | 74 | 16.3 | 28.6 | -0.42 | +0.0102 |
| 800 | 18:00 | 243.5 | 77 | 16.9 | 220.6 | 73 | 16.0 | 29.0 | -0.48 | +0.0114 |
| 1000 | 06:00 | 129.8 | 56 | 13.0 | 122.0 | 55 | 12.7 | 26.4 | -0.41 | +0.0108 |
| 1200 | 06:00 | 53.6 | 37 | 8.7 | 44.2 | 34 | 8.1 | 24.5 | -0.38 | +0.0111 |
| 1400 | 06:00 | 0.0 | 0 | 0.0 | 0.0 | 0 | 0.0 | 9.8 | -0.35 | +0.0113 |
| 1500 | 06:00 | 0.0 | 0 | 0.0 | 0.0 | 0 | 0.0 | 0.0 | -0.34 | +0.0114 |
| 1600 | 04:52 | 118.1 | 58 | 12.9 | 108.1 | 57 | 12.4 | 24.7 | 0.11 | -0.0039 |

**Table 3.** Closed form against a year of numerical propagation with J2–J6, the Sun and Moon from DE440 and a conical shadow. Hi-fi columns are for the inclination retuned so the node is stationary under the full force model; "drift" is what the J2-only inclination of equation (2) leaves, and Δi the correction that cancels it. Penumbra is the time per year in partial shadow (both edges of every pass); the longest pass is umbra only.

Three things stand out. First, the closed form is accurate. The propagation's umbra hours are 8–23 hours a year below it (0.1–0.26% of the year), the days with eclipse agree to within three days, the longest pass to within a minute, and the eclipse-free cases stay eclipse-free: at 1400 km, the edge of the island, the satellite spends 9.8 hours a year in partial shadow, losing the equivalent of 1.6 hours of full sunlight, and never enters the umbra. The umbra deficit is just the cone — 8 to 10 s per pass times a thousand passes — and if the penumbra is counted by the fraction of the Sun it hides, which is what a solar array sees, the two models agree to within about 8 hours a year, because the cylinder is in effect the umbra plus half the penumbra. Second, the J2-only inclination of equation (2) is within 0.013° of the right value; left uncorrected it lets the node drift by 0.3–0.6° a year on the dawn–dusk cases (0.1° on the 1600 km case away from the line), which is negligible against the 12–26° windows above 1500 km but is a decade's worth of the 3.7° window at 1400 km. Third, the small advantage of 06:00 over 18:00 in Section 8 does not survive: at 800 km the propagation finds the two lobes within two hours a year of each other (222.5 against 220.6 h), so the perihelion[G28] effect is real but cancelled. Re-running the 800 km pair with J2 alone restores the closed form's 9.5-hour advantage for 06:00; adding $J_3$–$J_6$ removes it (221.1 against 220.3 h), and the Sun and Moon then move the totals by only a couple of hours. The culprit is the pear-shaped $J_3$ term, which forces a small eccentricity on a sun-synchronous orbit whose orientation relative to the Sun differs between the two lobes.

## Against flying missions

Three solar observatories fly dawn–dusk sun-synchronous orbits for exactly the reason this paper studies, and their operators publish when the eclipses come. Table 4 sets those published seasons beside the model's prediction from nothing but the altitude[G1] and the side of the dawn–dusk line. The dates agree to within a few days and the longest pass to within about a minute — including the fact that the two 06:00 missions have their seasons around the December solstice and Hinode, on the 18:00 side, around the June solstice, which is the asymmetry of Section 8 seen in flight. The published seasons are for real, slightly elliptical orbits with their own node drift, so this is as close as a closed form can be expected to get.

| Mission | Orbit | Published eclipse season | This model |
|---|---|---|---|
| PROBA-2 (ESA, 2009–) | 725 km, i = 98.3°, LTAN 06:00 | about 80 days, November to January, eclipses up to 18 min per orbit | 11 Nov – 31 Jan, 80 days, up to 18.3 min |
| IRIS (NASA, 2013–) | 620 × 670 km, i = 97.9°, LTAN 06:00 | eclipse-free from the start of February to the end of October; 7–8 months of continuous observation | (645 km) 7 Nov – 3 Feb, 88 days, up to 19.8 min |
| Hinode (JAXA/NASA, 2006–) | 680 km, i = 98.1°, dawn–dusk, 18:00 side | early May to early August, nights up to 20 min per 98-min orbit; nine months of continuous sunlight | 8 May – 4 Aug, 87 days, up to 19.2 min |

**Table 4.** Published eclipse seasons of three dawn–dusk solar missions against the closed-form prediction for their altitude (dates for the J2000 year; in a given year they shift by a day or two). Sources: references 8–10.

## Against announced systems

Since late 2025 three organisations have published plans for large computing platforms in sun-synchronous orbit, chosen for exactly the continuous sunlight this paper is about. Google's Project

Suncatcher describes an 81-satellite cluster in a 650 km dawn–dusk orbit, with two prototypes planned for 2027 [12]. Starcloud's FCC application places up to 88,000 satellites in shells of up to 50 km between 600 and 850 km, sun-synchronous with a nominal 06:00 crossing time [13]. SpaceX's January 2026 FCC application for an "Orbital Data Center System" covers up to one million satellites in 50 km shells from 500 to 2000 km, in sun-synchronous and 30° inclinations, with representative sun-synchronous planes at 500, 1000 and 2000 km in its Schedule S — the altitudes used in Table 5 [14]. Starcloud's filing gives a nominal 06:00 crossing time and Google's paper a dawn–dusk orbit; SpaceX's filing gives no local time, and for its sun-synchronous shells we assume the dawn–dusk line, the only choice consistent with its stated purpose; where a paper or filing names no side, the 06:00 line of Table 2 is used. Table 5 runs each through the closed form, and every one of these orbits is a preset in the visualizer.

| System | Filed orbit | Eclipse (h/yr) | Days with eclipse | Longest pass (min) | Sunlit (%) |
|---|---|---|---|---|---|
| Google Project Suncatcher | 650 km, dawn–dusk SSO | 338 | 87 | 19.7 | 96.2 |
| Starcloud | 600–850 km, SSO shells, 06:00 crossing (as filed) | 379 (600 km) to 204 (850 km) | 92 to 68 | 20.7 to 16.0 | 95.7 to 97.7 |
| SpaceX Orbital Data Center System | 500–2000 km, SSO shells | 474 (500 km), 130 (1000 km), 0 from 1400 km | 102, 55, 0 | 22.7, 13.0, 0 | 94.6, 98.5, 100 |
| SpaceX Orbital Data Center System | 500–2000 km, 30° shells (not sun-synchronous) | 3160 (500 km) to about 2000 (2000 km) | 365 (342–355 at 2000 km) | 35.8 to 35.0 | 64 to 77 |

**Table 5.** Announced orbital-computing systems, as filed or published by October 2026, run through the closed form at the stated altitudes. Starcloud's filing states a nominal 06:00 crossing and Google's paper a dawn–dusk orbit; SpaceX's filing gives no local time, so dawn–dusk is assumed for its sun-synchronous shells; dawn–dusk rows without a stated side use the 06:00 line, as in Table 2. The 30° shells are not sun-synchronous: their node precesses through all orientations, so β sweeps ±53° every few weeks and the orbit is in shadow for roughly a third of every revolution; only at 2000 km does |β| briefly exceed the threshold, giving a couple of weeks a year without eclipse (figures for the 30° row are the range over node orientations). Sources: references 12–14.

The pattern is the paper's result seen from the other side. Every announced sun-synchronous shell below 1400 km has an eclipse season [G26] of two to three months, with passes of 13 to 23 minutes; Suncatcher at 650 km would spend about 340 hours a year in shadow, the same regime as the solar observatories in Table 4. Only the top of SpaceX's range reaches the eclipse-free island, and at 1400 km the window is 3.7° wide. SpaceX's narrative makes the case in the same terms as this paper — sunlight "up to more than 99% of the time for high-altitude, sun-synchronous orbits", with the lower-inclination shells load-balancing compute against time-variable demand [14]; the closed form puts that 99% line at about 1110 km for a dawn–dusk plane (98.5% sunlit at 1000 km, 99.4% at 1200 km) and 100% from 1400 km. The lower altitudes are not an oversight: debris-mitigation rules, the inner radiation belt above about 1000 km, launch mass and link latency all argue for flying low, and the 30° shells trade sunlight for cheaper launch and lower latency: at 500 km they are in shadow 36% of the time, about 3200 hours a year, in passes of up to 36 minutes. What the model adds is the price of each choice in hours of darkness, battery depth and thermal cycles — the quantities a power and thermal design is sized by.

Both implementations are open source. The Python model is published as the kelvinzero package [11] (`pip install kelvinzero`): a few hundred lines of NumPy organised exactly as this paper is, with the test suite, a command-line tool that prints the statistics for any orbit, a script that regenerates every figure here, and a self-contained 3-D viewer, so that every number in this paper can be regenerated and inspected locally.

## 11 Limitations, and what else to consider

The shadow is cylindrical: there is no penumbra[G24] and no atmospheric refraction, and the Sun is at infinity. With the real Sun–Earth distance the umbra is a cone of half-angle 0.26°, which at $\beta = 0$ shortens a pass by 8.4 s at 550 km, 8.9 s at 800 km and 10.2 s at 1500 km out of about 35 minutes, while the penumbra adds a partial fringe of 8.5–10.4 s at each end; Section 10 shows what that does to a year. Orbits are circular and propagated with the secular J2[G3] rate only; Section 10 supplies the full numerical cross-check (J2–J6, the Sun and Moon, a conical shadow), which moves eclipse hours by a few percent at most, and Orekit's propagators [7] are an independent way to repeat it. Earth is a sphere of equatorial radius. Natural extensions are the analemma[G14]-following node schedule, a Sun-centred inertial frame for the dynamics, Earth-fixed ground tracks with repeat cycles, and a conical umbra/penumbra term in the closed form itself. The model layer is framework-free and unit-tested, so each can be added with a reference value in the verification script.

### Eclipse is one problem among several

This paper isolates one question — can a fixed plane stay out of the shadow — and answers it. Choosing an orbit for a real mission weighs that answer against other things that the same altitude[G1] band changes, and it is worth naming them, without analysing them here:

1. **Radiation.** 1400–2000 km lies inside the inner Van Allen belt's proton population. Total dose and single-event rates there are far above those at 500–800 km, and this — more than any orbital subtlety — is why so little flies at these altitudes. Shielding, radiation-tolerant parts and detector design become first-order costs. There is precedent for the band: Globalstar has operated its constellation at 1414 km since 1998 and OneWeb flies at 1200 km.
2. **Launch and disposal.** Reaching 1500 km takes more Δv than 550 km, and natural decay there takes centuries, so end-of-life disposal — a lowering burn of a few hundred m/s, or a graveyard raise — is mandatory under 25-year and 5-year disposal rules rather than something the atmosphere does for free.
3. **Distance to the ground.** At 1500 km the satellite is about three times farther from its target than at 500 km: ground resolution coarsens for the same aperture and link margins fall, while swath, revisit and visibility time improve. For an optical imager the dawn–dusk[G9] ground track is lit at grazing angles; radar, communications, timing and astronomy payloads are unaffected.
4. **Keeping the node in the window.** Any error in inclination[G2] or altitude changes the precession[G4] rate and lets the node drift along the window (Section 10 measures the drift the J2-only inclination leaves, and the correction). At 1500–2000 km the windows are 12–26° wide and tolerate years of small drift; at the 1400 km edge the window is under 4°, so node maintenance or a higher altitude is needed.

5. **“Steady” thermal, not constant.** With no eclipse the solar input is constant, but Earth’s albedo and infrared emission still vary along the orbit and with season, and the β angle [G22] itself moves between about 54° and 90° over the year, so the Sun-facing geometry is steady rather than fixed.
6. **Rarer shadows.** The Moon occasionally occults the Sun as seen from the satellite (minutes, a few times a decade), and the shadow of the oblate Earth is slightly shorter in the polar direction than the spherical model assumes; neither changes the picture.
7. **Traffic.** The 1200–1500 km band already hosts large constellations; conjunction coordination and the debris environment there are part of the trade.

Acknowledgements and credits. The visualizer is built with React, three.js and KaTeX (MIT licences); its Earth imagery is NASA Blue Marble, public domain, via the example textures of three-globe (MIT). The model is NumPy; the high-fidelity check uses SciPy and jplephem, reading the JPL DE440 ephemeris (NASA/JPL). The typefaces are Source Serif 4, Inter and JetBrains Mono under the SIL Open Font License. The analemma photograph is by Giuseppe Donatiello (CC0) [3]. Full third-party notices: colure.space/kelvinzero/THIRD-PARTY-NOTICES.txt.

## 12 Glossary

The terms used in this paper, in the order a reader meets them. In the text, the first use of each term in a section carries a small G-number pointing here; figure references point to the picture that shows each one.

| | | |
|---|---|---|
| G1 | **Altitude, h** | Height of the circular orbit above Earth’s equatorial radius R = 6378.137 km. Sets the period, the sun-synchronous inclination and how easily the satellite sees over Earth’s edge (Figure 1, Figure 2a). |
| G2 | **Inclination, i** | Angle between the orbit plane and the equator. Sun-synchronous orbits are *retrograde*: i is a little more than 90°, so the satellite travels slightly against Earth’s rotation and the plane leans past the pole (Figure 2a). |
| G3 | **J2, the equatorial bulge** | The largest departure of Earth’s gravity from a perfect sphere: the equatorial radius is 21 km larger than the polar radius. Its pull slowly pivots an inclined orbit plane around the spin axis — *nodal precession*, equation (1). |
| G4 | **Nodal precession** | The slow rotation of the orbit plane about Earth’s spin axis caused by J2 (equation (1)): westward for orbits inclined less than 90°, eastward for retrograde ones. |
| G5 | **Sun-synchronous orbit (SSO)** | An orbit whose nodal precession equals the mean Sun’s eastward rate, 0.9856° per day, so the plane keeps a fixed angle to the mean Sun all year (equation (2), Figure 3). For each altitude there is exactly one such inclination (Table 1). |
| G6 | **Ascending node and RAAN** | The point where the orbit crosses the equator going north, and the angle Ω from a reference direction to it (right ascension of the ascending node). It fixes the orientation of the plane around the spin axis (Figure 2b). |
| G7 | **Node offset, $\Omega_{off}$** | The RAAN measured from the mean Sun instead of from the stars — constant for a sun-synchronous orbit. $\Omega_{off} = 0°$ is a noon–midnight plane, 90° and 270° are dawn–dusk planes. The visualizer's "RAAN offset from mean Sun" slider (Figure 2b). |
| G8 | **LTAN** | Local time of the ascending node: the local solar time on the ground beneath the satellite as it crosses the equator northbound, LTAN = 12:00 + $\Omega_{off}/15°$ (equation (3)). 15° of node offset is one hour. |
| G9 | **Dawn–dusk orbit** | A sun-synchronous plane with LTAN 06:00 or 18:00, i.e. $\Omega_{off} = 270°$ or 90°. Its normal points at the mean Sun, so the satellite flies along the day–night line and the plane faces the Sun; the natural candidate for an eclipse-free mission (Figure 19). |
| G10 | **Noon–midnight orbit** | A sun-synchronous plane with LTAN 12:00 or 00:00 ($\Omega_{off} = 0°$ or 180°). The plane contains the Sun direction, so the satellite crosses the full width of the shadow on every revolution (Figure 11). |
| G11 | **Mean Sun** | A fictitious Sun that moves eastward along the celestial equator at a perfectly uniform rate — the Sun we would have if Earth's orbit were circular and its axis upright. Clocks keep mean solar time; sun-synchronous orbits lock to the mean Sun (Section 3). |
| G12 | **Equation of time, EoT** | The difference between the real and the mean Sun along the equator, $EoT = L - \alpha$, up to ±16 minutes (±4°). Positive when a sundial runs ahead of the clock. It is the east–west width of the analemma (Figure 6b). |
| G13 | **Declination, $\delta$** | The Sun's latitude on the celestial sphere, from −23.44° at the December solstice to +23.44° at the June solstice. It is the north–south height of the analemma (Figure 6b). |
| G14 | **Analemma** | The figure-8 traced by the sub-solar point over a year in the mean-Sun frame: declination against minus the equation of time (Figure 6, Figure 8). In MSSI it is the set of directions from which sunlight reaches a frozen orbit. |
| G15 | **Ecliptic and obliquity, $\varepsilon$** | The plane of Earth's orbit around the Sun, and the 23.44° tilt of Earth's spin axis away from its perpendicular. The tilt gives the seasons and part of the equation of time (Figure 5). |
| G16 | **Vernal equinox direction** | The fixed direction among the stars where the Sun crosses the equator northbound around 20 March; the x axis of the ECI frame (Figure 3). |
| G17 | **ECI — Earth-Centred Inertial** | The standard inertial frame of orbital mechanics, in its J2000 (mean equator and equinox) realisation: origin at Earth's centre, z along the spin axis, x toward the vernal equinox, axes fixed to the stars. SSO planes precess and the Sun circles the ecliptic in it (Section 4, Figure 3). |
| G18 | **MSSI — Mean-Sun Synchronous Inertial** | ECI rotated about the spin axis by the mean solar longitude L(d), so x always points at the mean Sun (equation (4)). Sun-synchronous planes are frozen in it and the real Sun traces the analemma (Figure 3, Figure 4). |
| G19 | **Days since J2000, d** | The paper's time variable: days from 2000-01-01 12:00. The timeline covers one tropical year, 365.2422 days. |
| G20 | **Sub-solar point** | The point on Earth directly beneath the Sun; the Sun direction drawn on the globe (Figure 8). |
| G21 | **Terminator** | The day–night boundary: the great circle perpendicular to the Sun direction (Figure 11). |
| G22 | **$\beta$ angle** | The elevation of the Sun above the orbit plane, $\sin\beta = \hat{h} \cdot \hat{s}$. It is the one number that decides how much of each revolution is in shadow (Figure 9, Figure 10). |

G23 **Cylindrical shadow** — The shadow model used here: Earth's shadow is a cylinder of radius R pointing away from the Sun, with the Sun at infinity (no penumbra). A point is eclipsed if it is on the night side and within R of the Sun–Earth axis, equation (7) (Figure 9).

G24 **Umbra and penumbra** — The full shadow (umbra): a cone behind Earth where the Sun's disc is completely hidden; and the partial shadow around it (penumbra) where it is partly hidden. The cylinder used in this paper is the limit of the Sun at infinity; the difference is a few seconds per pass (Section 11) and is measured in Section 10.

G25 **Eclipse fraction, f_E** — The fraction of one revolution spent in shadow, a closed form in β and altitude (equation (8)), zero once |β| ≥ asin(R/(R+h)) — 67° at 550 km, 60° at 1000 km. Averaged over the year it gives the yearly eclipse (Figure 10, Figure 14).

G26 **Eclipse season** — The run of days around a solstice when a nearly-dawn–dusk orbit dips below its β threshold and is eclipsed on every revolution; about 74 days at 800 km, none above ~1400 km (Figure 14b, Figure 15).

G27 **Eclipse-time regions, U and I, α** — A map of every direction around Earth, in MSSI, coloured by what a point of the orbit shell there sees over the year: green — sunlit all year; orange — sunlit for part of the year, in shadow for the rest (between ∂U and ∂I); purple — in shadow all year. An orbit ring entirely in green is never eclipsed; one that enters orange has an eclipse season (eclipse-free part of the year, an eclipse on every revolution the rest); one that crosses purple is eclipsed on every revolution all year. The classification uses a cone of half-angle α = 90° + acos(R/(R+h)) around the Sun, equations (9)–(10) (Figure 12, Figure 13).

G28 **Perihelion and aphelion** — Earth's closest (early January) and farthest (early July) points from the Sun. Earth moves about 3% faster at perihelion, so the Sun's apparent drift is about 7% faster there; that variation is the eccentricity part of the equation of time and the reason the closed form gives 06:00 a slight edge over 18:00 — an edge the full force model takes back (Sections 8 and 10) (Figure 4).

## 13 References

1. U.S. Naval Observatory, Astronomical Applications Department, "Approximate Solar Coordinates", accessed 20 October 2026 — the low-precision solar formulae used in Section 5.
2. J. Meeus, *Astronomical Algorithms*, 2nd ed., Willmann-Bell, Richmond VA (1998), chapters 25 (solar coordinates) and 28 (equation of time) — the standard reference for the equation of time and the analemma.
3. G. Donatiello, "Analemma A14 2016", photograph, CC0 1.0 public-domain dedication, via Wikimedia Commons — image credit for Figure 7.
4. J. W. Spencer, "Fourier series representation of the position of the Sun," *Search* 2, 172 (1971) — the compact equation-of-time and declination series used as a cross-check.
5. D. A. Vallado, *Fundamentals of Astrodynamics and Applications*, 4th ed., Microcosm Press, Hawthorne CA (2013) — J2 nodal regression, the sun-synchronous condition, sidereal time.
6. J. R. Wertz and W. J. Larson (eds.), *Space Mission Analysis and Design*, 3rd ed., Microcosm Press / Kluwer (1999) — eclipse fraction of a circular orbit versus β angle.
7. L. Maisonobe, V. Pommier-Maurussane and P. Parraud, "Orekit: an open source library for operational flight dynamics applications," 4th International Conference on Astrodynamics Tools and Techniques (ICATT), ESA/ESAC, Madrid (2010); software: Orekit 12, accessed 20 October 2026 — an independent high-fidelity propagator, cited in Section 11; the cross-check of Section 10 uses the paper's own propagator (`kelvinzero.hifi`, reference 11).
8. S. Santandrea, K. Gantois, K. Strauch et al., "PROBA2: Mission and Spacecraft Overview," *Solar Physics* 286, 5–19 (2013), doi:10.1007/s11207-013-0289-5; and ESA PROBA2 Science Centre, "About PROBA2", accessed 20 October 2026 — the dawn–dusk orbit and its eclipse season (about 80 days from November to January, up to 18 minutes).

9. B. De Pontieu, A. M. Title, J. R. Lemen et al., "The Interface Region Imaging Spectrograph (IRIS)," *Solar Physics* 289, 2733–2779 (2014), doi:10.1007/s11207-014-0485-y; LMSAL, "How is IRIS operated," IRIS Technical Note 51; and eoPortal, "IRIS", accessed 20 October 2026 — 620 × 670 km, 97.9°, LTAN 06:00; eclipse-free from February to October.
10. T. Kosugi, K. Matsuzaki, T. Sakao et al., "The Hinode (Solar-B) Mission: An Overview," *Solar Physics* 243, 3–17 (2007), doi:10.1007/s11207-007-9014-6; eoPortal, "Solar-B / Hinode"; and the Hinode XRT Analysis Guide, accessed 20 October 2026 — 680 km, 98.1°, dawn–dusk; eclipse season from early May to early August with nights up to 20 minutes per 98-minute orbit.
11. K. Cooper and Y. Nikas, kelvinzero, software, MIT licence, doi:10.5281/zenodo.23382386 (concept DOI, resolving to the latest release) — the open-source Python implementation of this model (`pip install kelvinzero`); and the interactive visualizer at colure.space/kelvinzero.
12. Google Research, "Towards a future space-based, highly scalable AI infrastructure system design," technical paper (November 2025), summarised at research.google, accessed 20 October 2026 — Project Suncatcher: an illustrative 81-satellite cluster in a 650 km dawn–dusk sun-synchronous orbit; two prototype satellites with Planet planned for early 2027.
13. Starcloud, Inc., "Starcloud Orbital Datacenter System," FCC application SAT-LOA-20260202-00073, call sign S00803 (filed 4 February 2026; accepted for filing 13 March 2026), Narrative and Technical Attachment — up to 88,000 satellites in near-circular shells of up to 50 km between 600 and 850 km, in sun-synchronous inclinations with a nominal 06:00 crossing time; and Starcloud, "Why we should train AI in space," white paper, accessed 20 October 2026.
14. Space Exploration Holdings, LLC (SpaceX), "Orbital Data Center System," FCC application SAT-LOA-20260108-00016, Narrative, Technical Attachment and Schedule S (filed 30 January 2026; accepted for filing by FCC Public Notice DA 26-113, 4 February 2026) — up to one million satellites at 500–2000 km in shells of up to 50 km, in 30° and sun-synchronous inclinations, with representative sun-synchronous planes at 500, 1000 and 2000 km in the Schedule S; no local time of the node is stated.

Use of AI tools: Generative AI was used throughout this work, including to draft and edit the text, to write the model, test, propagation and figure code, and to build the visualizer. The problem, the mean-Sun frame, the design of the verification and every claim are the work of the authors. The authors checked every derivation, number and reference, verified the closed form independently against a full-force propagation and three flying missions, and take full responsibility for the content of this work.

Constants: $\mu = 398600.4418$ km$^3$/s$^2$, $R = 6378.137$ km, $J_2 = 1.08262668 \times 10^{-3}$, tropical year 365.2422 d, mean-Sun rate 0.9856474°/day, obliquity 23.439°, $e = 0.0167086$, J2000.0 = 2000-01-01 12:00.

## GET IN TOUCH

Questions, corrections, or a mission that could use an eclipse-free orbit: write to hello@colure.space, or open a thread in the discussions on GitHub, where the technical conversation lives.

Kyle Cooper on LinkedIn ↗ Yanni Nikas on LinkedIn ↗